\documentclass[journal,10pt]{IEEEtran}

\usepackage{graphicx, enumerate, amsmath, yhmath, amssymb, psfrag,algorithm, algorithmic,xspace,tikz,cancel}
\usepackage[caption=false,font=footnotesize]{subfig}
\usetikzlibrary{shapes,arrows,dsp,chains}
\tikzstyle{line} = [draw, -latex']

\makeatletter
\newcommand\newsubcommand[3]{\newcommand#1{#2\sc@sub{#3}}}
\def\sc@sub#1{\def\sc@thesub{#1}\@ifnextchar_{\sc@mergesubs}{_{\sc@thesub}}}
\def\sc@mergesubs_#1{_{\sc@thesub_#1}}
\makeatother

\newcommand{\dipp}{\ensuremath{\textsc{dipp}}\xspace}

\newcommand{\cs}{\textsc{cs}\xspace}
\newcommand{\dcs}{\textsc{dcs}\xspace}

\newcommand{\consensus}{\ensuremath{\textup{\texttt{consensus}}}\xspace}
\newcommand{\majority}{\ensuremath{\textup{\texttt{majority}}}\xspace}
\renewcommand{\sp}{\ensuremath{\textsc{sp}}\xspace}
\newcommand{\cosamp}{\ensuremath{\textsc{c}\text{o}\textsc{s}\text{a}\textsc{mp}}\xspace}

\newcommand{\mbx}{\mathbf{x}}
\newcommand{\mbA}{\mathbf{A}}
\newcommand{\mby}{\mathbf{y}}
\newcommand{\mbe}{\mathbf{e}}

\renewcommand{\complement}{\ensuremath{\mathsf{c}}}

\newtheorem{definition}{Definition}
\newtheorem{example}{Example}

\newtheorem{proposition}{Proposition}
\newtheorem{remark}{Remark}
\newtheorem{corollary}{Corollary}
\newtheorem{lemma}{Lemma}

\newtheorem{assumption}{Assumption}

\newcommand{\wqed}{\hfill \ensuremath{\square}}
\newcommand{\addi}{\texttt{vote}_1}

\newcommand{\T}{\mathcal{T}}
\newcommand{\Th}{\hat{\mathcal{T}}}

\newcommand{\I}{\mathcal{I}}
\newcommand{\p}{\ensuremath{\mathsf{p}}}
\newcommand{\q}{\ensuremath{\mathsf{q}}}
\renewcommand{\r}{\ensuremath{\mathsf{r}}}
\newsubcommand{\Ti}{\ensuremath{\mathcal{T}}}{\p}
\newsubcommand{\Tii}{\ensuremath{\mathcal{T}}}{\q}
\newsubcommand{\Tiii}{\ensuremath{\mathcal{T}}}{\r}

\newsubcommand{\Thi}{\ensuremath{\hat{\mathcal{T}}}}{\p}
\newsubcommand{\Thii}{\ensuremath{\hat{\mathcal{T}}}}{\q}
\newsubcommand{\Thiii}{\ensuremath{\hat{\mathcal{T}}}}{\r}

\newcommand{\J}{\ensuremath{\mathcal{J}}}
\newcommand{\Jh}{\ensuremath{\hat{\mathcal{J}}}}
\newsubcommand{\Ii}{\ensuremath{\mathcal{I}}}{\p}
\newsubcommand{\Iii}{\ensuremath{\mathcal{I}}}{\q}
\newsubcommand{\Iiii}{\ensuremath{\mathcal{I}}}{\r}

\newcommand{\gp}{\textsc{gp}\xspace}

\newcommand{\smnr}{\textsc{smnr}\xspace}

\renewcommand{\appendixname}{appendix\xspace}

\newcommand{\algorithmname}{algorithm}
\newcommand{\propositionname}{proposition}

\renewcommand{\figurename}{Fig.}

\renewcommand{\P}[1]{\ensuremath{\mathbb{P}\,\!\bigl(#1\bigr)}}
\newcommand{\sP}[1]{\ensuremath{\mathbb{P}\,\!(#1)}}

\ifCLASSINFOpdf
\else
\fi
\allowdisplaybreaks
\begin{document}
%\title{Hard Support-set Detection by Probabilistic Performance Modeling}
\title{Hard Decisions by Probabilistic Modeling}
\title{Analysis of Voting Principles for Support-set Estimation used in Greedy Pursuits}
\title{Analysis of Voting Principles for Support-set Estimation used in Distributed Greedy Pursuits}
\title{Analysis of Democratic Voting Principles used in Distributed Greedy Algorithms}
%\title{A Probabilistic Model for Hard Support Detection \\ used in  Distributed Parallel Pursuit}

\author{Dennis~Sundman,~\IEEEmembership{Student member,~IEEE,}
        Saikat~Chatterjee,~\IEEEmembership{Member,~IEEE,}
        and~Mikael~Skoglund,~\IEEEmembership{Senior~Member,~IEEE}% <-this % stops a space
\thanks{The authors are with Communication Theory Laboratory, School of Electrical Engineering, KTH - Royal Institute of Technology, Sweden. 
Emails: $\{$denniss, sach, skoglund$\}$@kth.se}}
%\thanks{}% <-this % stops a space
%\thanks{Manuscript received April 19, 2005; revised December 27, 2012.}}


\markboth{Journal of \LaTeX\ Class Files,~Vol.~11, No.~4, December~2012}%
{Shell \MakeLowercase{\textit{et al.}}: Bare Demo of IEEEtran.cls for Journals}
% after the title page when using the twoside option.%
 



\maketitle

\begin{abstract}
  A key aspect for any greedy pursuit algorithm used in compressed sensing is a good support-set detection method. For distributed compressed sensing, we consider a setup where many sensors measure sparse signals that are correlated via the existence of a signals' intersection support-set. This intersection support-set is called the joint support-set. Estimation of the joint support-set has a high impact on the performance of a distributed greedy pursuit algorithm. This estimation can be achieved by exchanging local support-set estimates followed by a (consensus) voting method. In this paper we endeavor for a probabilistic analysis of two democratic voting principle that we call majority and consensus voting. In our analysis, we first model the input/output relation of a greedy algorithm (executed locally in a sensor) by a single parameter known as probability of miss. Based on this model, we analyze the voting principles and prove that the democratic voting principle has a merit to detect the joint support-set.
\end{abstract}

% \begin{abstract}
% One key ingredient in any greedy pursuit algorithm for compressed sensing is good support-set detection; knowing the support, signal recovery is straight forward with linear least squares. In a multiple sensor-node environment where sensors measure sparse signals with partial joint support-sets, improved support-set estimation accuracy can be achieved by firstly exchanging local support-set estimates; and secondly merge the knowledge from the received support-sets to achieve a consensus estimate. In this paper we introduce a single parameter linear probabilistic framework for modeling the performance of local single-sensor greedy pursuit algorithms. This model is then applied in two scenarios to show the suitability of democratic voting as consensus method for support-detection in a multiple-node environment.
% \end{abstract}

\begin{IEEEkeywords}
Greedy algorithms, distributed detection, hard decision.
\end{IEEEkeywords}

% \fi
\IEEEpeerreviewmaketitle

\section{Introduction}
\IEEEPARstart{C}{}{ompressed} sensing (\cs)~\cite{Donoho:compressed_sensing,Candes:stable_signal_recovery} typically considers a single-sensor scenario, where the main task is reconstruction of a large-dimensional signal-vector from a small-dimensional measurement-vector by using a-priori knowledge that the signal is sparse in a known domain. 
Several \cs reconstruction algorithms have been considered in the literature, for example convex optimization- \cite{Mota:distributed_basis_pursuit, Bazerque:distributed_spectrum_sensing}, Bayesian- \cite{Ji:bayesian_compressive_sensing, Donoho:message_passing_for_cs} and greedy pursuit (\gp) algorithms.
The greedy pursuit (\gp) algorithms are popular due to their low complexity and good performance. From a measurement vector, the \gp algorithms use linear algebraic tools to estimate the underlying \textit{support-set} of the sparse signal-vector followed by estimating associated signal values; here we mention that good support-set estimation is a key aspect for the \gp algorithms.
A few examples of typical \gp algorithms include: matching pursuit \cite{Mallat:matching_pursuit_with_time_frequency_dictionaries}, orthogonal matching pursuit (\textsc{omp})~\cite{Tropp:signal_recovery}, \cosamp \cite{Needell:cosamp}, subspace pursuit (\sp)~\cite{Dai:subspace_pursuit}, but there are many others~\cite{Donoho:sparse_solution_of_underdetermined_systems_stomp,Chatterjee:projection_based_look_ahead,Sundman:frogs,Needell:signal_recovery_from_incomplete_and_inaccurate_measurements_via_romp,Sundman:look_ahead_parallel_pursuit}. For the \gp algorithms, just as for any \cs reconstruction algorithm, providing analytical performance guarantees is an important yet challenging task. These guarantees are typically done through worst case analysis based on restricted isometry property~\cite{Candes:restricted_isometry_property} and mutual coherence inequalities.

Distributed (or de-centralized) \cs (\dcs) refers to a problem of multiple sensors connected over a network, where the sensors observe correlated sparse signals through \cs measurements.
By the term \dcs we refer both to simultaneous estimation in a distributed network~\cite{Tropp:simultaneous_sparse_approx_part1,Rakotomamonjy:surveying,Leviatan:simultaneous,Cotter:sparse_solutions,Chen:theoretical_results_on_sparse,Sundman:greedy_pursuit_for_jointly} and to multiple measurement vector setups in some fusion center~\cite{Mota:distributed_basis_pursuit,Bazeraque:distributed_spectrum_sensing,Feng:distributed_compressive_spectrum_sensing,Ling:decenteralized_support_detection}.
Recently we developed several \gp algorithms for \dcs, called distributed greedy pursuits~\cite{Sundman:diprsp,Sundman:a_greedy_pursuit_algorithm,Sundman:distributed_gp_algorithms, Sundman:dipp_arxiv}.
%We have recently published distributed \dcs algorithms based on \gp pursuits  and most recently [DIPP]. In this paper it is typically not necessary to specifically distinguish between a distributed or a centralized solver.
%
In \dcs, two (of many) models for signal correlations are the common support-set model and the mixed support-set model~\cite{Sundman:methods_for_dcs}. In the common support-set model, the same (joint) \emph{full} support-set is assumed for all signals measured at different sensors, while in the mixed support-set model a joint \emph{partial} support-set is assumed for all sensors. Based on these models, a prominent approach for distributed \gp algorithms is to let the sensors in the network exchange (or transmit to a centralized point) full support-set estimates and then, using only support-set knowledge, estimate the joint support-set. A better estimate of the joint support-set can then be used to improve \dcs reconstruction performance.

In general, theoretical performance analysis of distributed \gp algorithms is non-trivial and we recently developed \dipp (distributed parallel pursuit) -- a distributed greedy pursuit algorithm -- with such theoretical guarantees in~\cite{Sundman:dipp_arxiv}. Through analysis and simulations we have shown that \dipp performs better than local \gp algorithms, such as \sp. In \dipp and other distributed greedy pursuits, the joint support-set is estimated by a consensus voting method. In several of our earlier works~\cite{Sundman:a_greedy_pursuit_algorithm,Sundman:distributed_gp_algorithms}, we assumed that democratic based voting is suitable for consensus, and in \cite{Sundman:dipp_arxiv} we proved theoretical reconstruction guarantees based on this assumption. The advantage of voting has earlier been studied in politics and finance as early as 1785 \cite{Young:optimal_voting_rules,Ledyard:the_approximation_of_efficient}. In this paper, we endeavor to prove that the assumption of democratic voting for support-set estimation, based on \gp algorithms, indeed has a merit. In our approach, we assume that support-sets estimated from \gp algorithms executed locally in 
several sensors likely contain independent errors. Therefore, based on probability of detection, miss, and false alarms, we first model the input/output relation of relevant \gp algorithms by using standard detection theory framework. 

Using the input/output relation, we provide probability results for consensus strategies based on democratic voting applied in scenarios that employ the common and mixed support-set models. The main contributions of this paper can be summarized as:
\begin{itemize}
\item Defining the input/output relation of relevant \gp algorithms.
\item Probabilistic analysis of democratic voting used in distributed \gp algorithms for both common and mixed support-set models.
\end{itemize}

The outline of the paper is as follows: We first introduce some notation in the next subsection. Then, in Section~\ref{sec:system_model}, we introduce the signal model, the common support-set, and mixed support-set models. In Section~\ref{sec:prob_model}, we introduce an input/output relation to model single sensor performance, which is then used for analyzing different voting strategies for the the common support-set model in Section~\ref{sec:common_model} and the mixed support-set model in Section~\ref{sec:mixed_model}. Then, in Section~\ref{sec:experiment}, we provide experimental verification of the results achieved.

\subsection{Notation}
Sets are denoted by calligraphic capital letters, in particular; $\T$, $\I$ and $\J$ are support-sets or partial support-sets. We define the full set $\Omega = \{ 1,2, \dots, N \}$ and the set complement $\J^{\complement} = \Omega \setminus \J$, where `$\setminus$' is the set-minus. We denote the event of an index $i$ residing in the support-set $\T$ by $i \in \T$. If $i$ resides in two support-sets -- $\Ti$ and $\Tii$ -- we use either $(i\in \Ti, i\in\Tii)$ or $i\in (\Ti \cap \Tii)$; where the one providing most insight will be used. The probability of an index $i$ residing inside the support-set $\T$ is denoted by $\P{i\in\T}$. Lastly, we denote the conditional probability, where by $\P{i\in\J|i\in\T}$ we refer to the probability that an index $i$ be in $\J$, given that $i$ is (randomly) in $\T$. Lastly we introduce two algorithmic notations
\begin{align} 
\addi(\mathbf{z}, \T) & \triangleq \{ \forall i \in \T, \textit{ perform } z_{i} = z_{i}+1 \}.
\end{align}
\begin{align}
  \texttt{max\_indices}(\mathbf{z}, T) \triangleq & \{\textit{select the $T$ largest amplitude} \nonumber \\
    & \,\,\, \textit{indices of $\mathbf{z}$} \}.
\end{align}
%where $\mathbf{z}=[z_{1} \,\, z_{2} \,\, \ldots z_{N} ]$ and $z_{j} \geq 0$.

\section{System Model}\label{sec:system_model}
In this section we define the distributed compressed sensing (\dcs) problem, the common support-set model and the mixed support-set model.

\subsection{Distributed Compressed Sensing}
In distributed compressed sensing (\dcs), each $\p$'th sensor measures a signal $\mathbf{x}_{\p} \in \mathbb{R}^{N}$ through the following linear relation
\begin{align}
\mathbf{y}_{\p} = \mathbf{A}_{\p} \mathbf{x}_{\p} + \mathbf{e}_{\p}, ~~~~~~ \forall \p \in \mathcal{L}, \label{eqn:model}
\end{align}
where $\mathbf{y}_{\p} \in \mathbb{R}^{M}$ is a measurement vector, $\mathbf{A}_{\p} \in \mathbb{R}^{M\times N}$ is a measurement matrix, $\mathbf{e}_{\p} \in \mathbb{R}^{M}$ is some measurement noise and $\mathcal{L}$ is a global set containing all sensors (nodes) in the network ($|\mathcal{L}| = L$). The signal vector $\mathbf{x}_{\p} = [x_{\p}(1) \,\, x_{\p}(2) \, \ldots \, x_{\p}(N)]$ is $T$-sparse, meaning it has $T$ elements that are non-zero. Thus, the setup describes an under-determined system, where $T < M < N$. The element-indices corresponding to non-zero values are collected in the support-set $\Ti$; that means $\Ti = \{ i : x_{\p}(i) \neq 0 \}$ and $| \Ti | = T$. A dense vector containing only the non-zero values of $\mbx_{\p}$ is represented by $\mathbf{v}_{\p} = [x_{\p}(\Ti(1)), x_{\p}(\Ti(2)), \dots, x_{\p}(\Ti(T))]$, which may also be independent locally and across the network. Throughout this paper we use measurement matrices that have unit $\ell_2$-norm columns and characterize the signal-
to-noise ratio using the signal-to-measurement-noise-ratio (\smnr), which is defined for sensor $\p$ as
\begin{align}
  \smnr \triangleq \frac{ \mathcal{E} \{ \| \mathbf{x}_{\p} \|_{2}^{2} \} }{ \mathcal{E} \{ \| \mathbf{e}_{\p} \|_{2}^{2} \} }. \label{eqn:smnr}
\end{align}
Furthermore, $\mathbf{A}_{\p}$ and $\mathbf{e}_{\p}$ are independent both locally and across the network.

In order to benefit from cooperation in the network, some correlation in the signal vector $\mbx_{\p}$ must be present. In the following two subsections we present these correlations by introducing the common and mixed support-set models.

\subsection{Common Support-set Model}\label{sec:common_support_signal_model}
In the common support-set model~\cite{Duarte:distributed_compressed_sensing,Sundman:methods_for_dcs}, the support-sets of all signals in the network $\mbx_{\p}$ are identical. That is
\begin{align}
\T_{\p} = \J ~~~~~ \forall \p \in \mathcal{L}, \label{eqn:common}
\end{align}
where we refer to $\J$ as the \emph{joint} support-set.

\subsection{Mixed Support-set Model}\label{sec:mixed_support_signal_model}
A natural extension to the common support-set model is the mixed support-set model, proposed by us in~\cite{Sundman:greedy_pursuit_for_jointly, Sundman:a_greedy_pursuit_algorithm, Sundman:methods_for_dcs}. In this case there exists an intersection between all support-sets $\T_{\p}$. Denoting $\J = \cap_{\forall \p \in \mathcal{L}} \T_{\p}$, we have
\begin{equation}
\T_{\p} = \I_{\p} \cup \J ~~~~~ \forall \p \in \mathcal{L}. \label{eqn:mixed}
\end{equation}
Here, we refer to $\J$ as the joint part of the support-set (or simply joint support-set) and $\I_{\p} \triangleq \T_{\p} \setminus \J$ is the individual part.
\begin{assumption} Denoting $|\Ii|=I ~ \forall ~ \p$ and $|\J|=J$, the following assumptions are used throughout the paper:
\begin{enumerate}
 \item Elements of support-sets are uniformly distributed,
   \begin{align}
     \P{i \in \Ti} = \frac{|\Ti|}{|\Omega|} = \frac{T}{N}, ~~ \forall \p\in \mathcal{L}. \label{eqn:cons:uniformity}
   \end{align}  
 \item $\Ii \cap  \J = \emptyset, ~~~ \forall \p\in \mathcal{L}$.
 \item $\Ii \cap \Iii = \emptyset, ~~~ \forall \p, \q \in \mathcal{L}, \p \neq \q$.
 \item Hence, $T = I+J$.\wqed
\end{enumerate}
\end{assumption}

\section{Modeling the input/output Relation for Greedy Pursuits}\label{sec:prob_model}
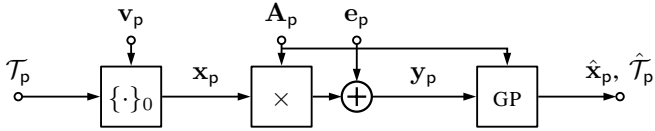
\begin{figure}[t]
  \centering
  \begin{tikzpicture}[node distance=1.5cm]
    \node[dspsquare] (nature) {$\{ \cdot \}_0$};
    \node[dspsquare,right of=nature,node distance=2cm] (mix1) {$\times$};
    \node[dspadder,right of=mix1,node distance=1cm] (add1) {};
    \node[dspsquare,right of=add1,node distance=2cm] (cs) {\gp};
    \node[dspnodeopen,dsp/label=above,right of=cs] (out) {$\hat{\mbx}_{\p}$, $\Th_{\p}$};
    \node[dspnodeopen,dsp/label=above,above of=add1,node distance=0.75cm] (e) {$\mbe_{\p}$};
    \node[dspnodeopen,dsp/label=above,left of=nature,node distance=1.5cm] (in) {$\Ti$};
    \node[dspnodeopen,dsp/label=above,above of=nature,node distance=0.75cm] (v) {$\mathbf{v}_{\p}$};
    \node[dspnodeopen,dsp/label=above,above of=mix1,node distance=0.75cm] (A) {$\mbA_{\p}$};
    \coordinate[above of=cs, node distance=0.64cm] (acs) {};
    \coordinate[above of=mix1, node distance=0.64cm] (aA) {};
    \draw[dspconn] (v) -- (nature);
    \draw[dspconn] (A) -- (mix1);
    \draw[dspconn] (aA) -- (acs) -- (cs.north);
    \draw[dspconn] (in) -- (nature);
    \draw[dspconn] (e) -- (add1);
    \draw[dspconn] (nature) -- node[midway,label={above:$\mbx_{\p}$}] {} (mix1);
    \draw[dspconn] (mix1) -- (add1);
    \draw[dspconn] (add1) -- node[midway,label={above:$\mby_{\p}$}] {} (cs);
    \draw[dspconn] (cs) -- (out);
  \end{tikzpicture}
  \caption{The \cs system considered in this paper. From known underlying support-set to estimate provided by reconstruction algorithm.}\label{fig:full_system}
\end{figure}

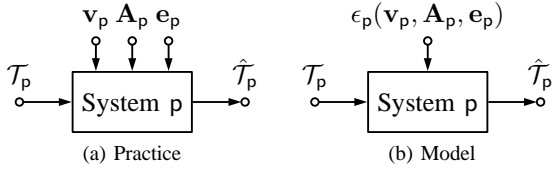
\begin{figure}[t]
  \centering
  \subfloat[Practice]{
    \begin{tikzpicture}[node distance=1.5cm]
      \node[dspsquare] (system) {~System \p~};
      \node[dspnodeopen,dsp/label=above,left of=system] (in) {$\Ti$};
      \node[dspnodeopen,dsp/label=above,right of=system] (out) {$\Thi$};
      \node[dspnodeopen,dsp/label=above] (v) at ([yshift=4mm] 
      $(system.north west)!0.2!(system.north east)$) {$\mathbf{v}_{\p}$};

      \node[dspnodeopen,dsp/label=above] (A) at ([yshift=4mm]
      $(system.north west)!0.5!(system.north east)$) {$\mbA_{\p}$};

      \node[dspnodeopen,dsp/label=above] (e) at ([yshift=4mm]
      $(system.north west)!0.8!(system.north east)$) {$\mbe_{\p}$};
      
      \draw[dspconn] (v) -- ($(system.north west)!0.2!(system.north east)$);
      \draw[dspconn] (A) -- ($(system.north west)!0.5!(system.north east)$);
      \draw[dspconn] (e) -- ($(system.north west)!0.8!(system.north east)$);
      \draw[dspconn] (in) -- (system);
      \draw[dspconn] (system) -- (out);
    \end{tikzpicture} \label{fig:system1}
  } \qquad
  \subfloat[Model]{\begin{tikzpicture}[node distance=1.5cm]
      \node[dspsquare] (system) {~System \p~};
      \node[dspnodeopen,dsp/label=above,left of=system] (in) {$\Ti$};
      \node[dspnodeopen,dsp/label=above,right of=system] (out) {$\Thi$};
      \node[dspnodeopen,dsp/label=above] (epsilon) at ([yshift=4mm] $(system.north west)!0.5!(system.north east)$) {$\epsilon_{\p}(\mathbf{v}_{\p}, \mathbf{A}_{\p}, \mbe_{\p})$};
      \draw[dspconn] (epsilon) -- ($(system.north west)!0.5!(system.north east)$);
      \draw[dspconn] (in) -- (system);
      \draw[dspconn] (system) -- (out);
    \end{tikzpicture} \label{fig:system2}
    } \caption{Two simplified figures of the full system.}
\end{figure}

A \gp algorithm in sensor $\p$ will attempt to find the true support-set $\Ti$. Influencing the chances of success are a number of factors: signal amplitudes (i.e., $\mathbf{v}_{\p}$), measurement noise $\mbe_{\p}$, sparsity $T$ and measurement matrix $\mathbf{A}_{\p}$ realization. Illustrated in \figurename~\ref{fig:full_system}, is the whole procedure from an underlying $\Ti$, signal acquisition according to \eqref{eqn:model}, to recovered support-set estimate $\Thi$. In \figurename~\ref{fig:system1}, we have simplified the previous figure in one box, referred to as the \textit{System}. Borrowing terms from detection theory we model the system (see Definition~\ref{ass}), where the idea is to replace the factors influencing support-set recovery performance with one single parameter $\epsilon_{\p}$, shown in \figurename~\ref{fig:system2}. Introduction of this single parameter helps to bring analytical tractability, which we will witness in Sections~\ref{sec:common_model} and \ref{sec:mixed_model}.

\begin{definition}[System model] \label{ass}
  The support-set estimate $\Thi$ of any unbiased \gp algorithm described in \figurename~\ref{fig:system2} follows
  \begin{align}
    \P{i \in \Thi} & & & = \frac{T}{N} \label{ass1} \\
    \P{i \in \Thi| i\in \Ti} & = \P{\text{``detect''}} & & = 1 - \epsilon_{\p} \label{ass2} \\
    \P{i \notin \Thi| i\in \Ti} & = \P{\text{``miss''}} && = \epsilon_{\p} \label{ass3} \\
    \P{i \in \Thi| i\in \Ti^{\complement}} & = \P{\text{``false alarm''}} & & = \frac{T}{N-T}\epsilon_{\p}\label{ass4},
  \end{align}
  where $0 \leq \epsilon_{\p} \leq \frac{N-T}{N}$. Observe that $i \notin \Thi = i\in \Thi^{\complement}$. These probabilities should be read as, for example in \eqref{ass2}: \textit{``The probability that index $i$ is part of the output $\Thi$ from the system, provided that this index is already part of the true underlying support-set $\Ti$''.} For the remainder of the paper, we assume that all sensors in the network have statistically identical system and signal conditions, meaning that $\epsilon_{\p} = \epsilon ~ \forall ~ \p$. \wqed

\textit{Discussion:} The input/output relation in Definition~\ref{ass} follows from symmetry arguments. Since the system is symmetric and $\Ti$ is uniformly random, any unbiased (fair) reconstruction algorithm will produce $\Thi$ which is also uniformly random \eqref{ass1}; an unbiased algorithm should not favor any correct index over than any other correct index, resulting in \eqref{ass2}. Similarly, the algorithm should not favor any missed index over another missed index \eqref{ass3}. Furthermore, whenever a support-index is missed, a false alarm has occurred; therefore the false alarm can be parametrized by the same parameter as the probability of miss and detect \eqref{ass4}. Using $\epsilon$ to specify the system behavior, we see that the worst possible system will select indices for the support-set uniformly at random. Thus the upper-bound on $\epsilon$ is $\epsilon_{\max} = \frac{N-T}{N}$, which means that the worst $\P{\text{``detect''}} = \frac{T}{N}$. \wqed
\end{definition}

At this point, there is no closed-form expression of the parameter $\epsilon$ as it would require complete characterization of $\mbA_{\p}$, $\mbx_{\p}$, $\mbe$ and of the present \gp algorithm. Such an analysis is outside the scope of this paper; instead we estimate $\epsilon$ through experiments. This can in practice be performed, for example, by using pilot signals. We now present the first result.
\begin{proposition} \label{prop:disconnect} The probability that an index `$i$' is correct for sensor $\p$ provided that it is found by the $\p$'th system is given by
  \begin{align}
    \P{i\in\Ti | i\in\Thi} = \P{i\in\Thi | i\in\Ti}. \label{eqn:disconnect}
  \end{align}
  \begin{IEEEproof}
    \begin{align}
      \P{i\in\Ti | i\in\Thi} & \overset{(a)}{=} \frac{\P{i\in\Thi | i\in\Ti}\P{i\in\Ti}}{\P{i\in\Thi}} \nonumber \\
      & \overset{(b)}{=} \frac{\P{i\in\Thi | i\in\Ti}\frac{T}{N}}{\frac{T}{N}} \nonumber \\
      & = \P{i\in\Thi | i\in\Ti}, \nonumber
    \end{align}
    where we in \ensuremath{(a)} have used Bayes' rule and in \ensuremath{(b)} have used \eqref{eqn:cons:uniformity} and \eqref{ass1}.
  \end{IEEEproof}
\end{proposition}

\subsection{Numerical Verification of the System Model} \label{sec:evaluation_prob_model}
In order to verify the system model in Definition~\ref{ass}, we perform two different tests. As \gp algorithm we have used the well known subspace pursuit (\sp) algorithm, however; similar results can be obtained with other \gp algorithms that are based on fixed support-set size.

\begin{figure}[t]
  \centering
  \psfrag{aaaaaaaa}{\footnotesize $\sP{i \in \Ti}$}
  \psfrag{bbb}{\footnotesize $T/N$}
  \includegraphics[width=\columnwidth]{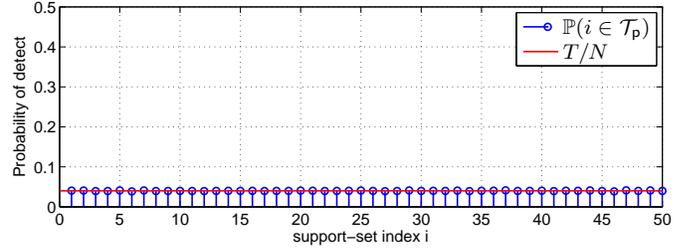}
  \caption{Figure showing how often each index occurs in the output of the system, based on \eqref{ass1} of the system model.} \label{fig:ass1}
\end{figure}

In the first test, presented in \figurename~\ref{fig:ass1}, we verify \eqref{ass1}. The test is based on $10^5$ random: support-sets $\Ti$, signal realizations $\mathbf{v}_{\p}$, measurement matrices $\mathbf{A}_{\p}$ and noises $\mathbf{e}_{\p}$ (generated such that $\smnr = 20$ dB). In \figurename~\ref{fig:ass1}, $N = 50$ and $T = 2$ to make the outcome observable ($M = 7$). Along the x-axis we show the support-set index $i$, and on the y-axis, we show how many times each index appears in the output $\Thi$. From this figure, we see that by using the proposed setup, the output from the algorithm is uniform, which verifies \eqref{ass1} of the definition.

\begin{figure}[t]
  \centering
  \psfrag{aaaaaaaaaaaaaaaaa}{\scriptsize $\sP{i \in \Thi | i \in [14, 26]}$}
  \psfrag{bbb}{\scriptsize $1-\hat{\epsilon}$}
  \includegraphics[width=\columnwidth]{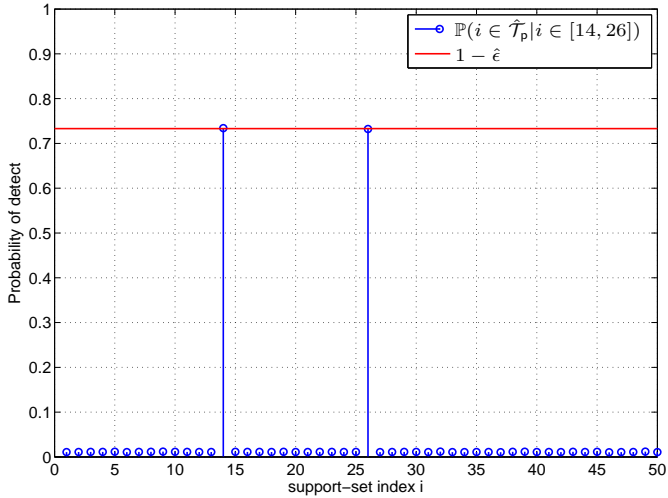}
  \caption{Simulation verification that indeed there exist an underlying $\epsilon$ such that the proposed system model holds.} \label{fig:ass2}
\end{figure}

In the second test, presented in \figurename~\ref{fig:ass2}; \eqref{ass2}, \eqref{ass3} and \eqref{ass4} are verified where $N = 50$ and $T = 2$ (and $M = 7$). Here, there are $10^5$ random: signal realizations $\mathbf{v}_{\p}$, measurement matrices $\mbA_{\p}$ and noises $\mathbf{e}_{\p}$ (such that $\smnr = 20$ dB). The support-set $\Ti = [14, 26]$ is fixed in order to produce an informative figure. Along the x-axis is the support-set index $i$, and on the y-axis, we show how many times each index appears in the output $\Thi$. We can now directly identify the three equations \eqref{ass2}, \eqref{ass3} and \eqref{ass4}. First, we estimate $\epsilon$ by counting the number of false alarms \eqref{ass4}; in this case $\hat{\epsilon} = 0.267$. Then \eqref{ass2} and \eqref{ass3} are found directly from $\hat{\epsilon}$. We will now apply the input/output relation model to more complex scenarios.

\section{Voting Based Detection for \\ the Common Support-set Model} \label{sec:common_model}
In this section we introduce the concept of voting based on support-set estimates from a number of nodes. We model signal correlation according to the common support-set model (see Section~\ref{sec:common_support_signal_model}). Throughout this section we use $\T$ and $\J$ interchangeably for the same purpose, since they are equivalent in the common support-set model.

Consider a setup with multiple sensor nodes where each sensor gathers \cs measurements and runs a local \gp algorithm to find a local support-set estimate. The support-set estimates from all nodes are then sent to a fusion center (or exchanged distributively) for estimation of $\J$.

\subsection{Algorithm} \label{sec:common_algo}
We propose a fusion center strategy based on democratic voting where, assuming $T$ is known, the strategy for the final estimate is to choose the $T$ indices with most votes. This is a \majority voting strategy and a detailed description is presented in Algorithm~\ref{alg:majority}.
\begin{algorithm}[ht] 
  \caption{\majority: %\newline
    \textit{Executed in the local node $\p$}} \label{alg:majority} 
  \textit{Input:} $\{ \hat{\T}_{\p} \}_{\p \in \mathcal{L}}$, $T$ \\
  \textit{Initialization:} $\mathbf{z} \leftarrow \mathbf{0}_{N\times 1}$ \label{alg:majority:cons:0} \\
  \textit{Algorithm:}
  \begin{algorithmic}[1]
    \FOR { each $\p \in \mathcal{L}$}
    \STATE $\mathbf{z} \leftarrow \addi(\mathbf{z}, \hat{\T}_{\p})$
    \hfill (The estimate of sensor $\p$) \label{alg:majority:cons:others}
    \ENDFOR
    % 
%    \STATE Choose $\hat{\J}$ \hfill such that \hfill $z(i) \geq z(j) ~ \forall i,j \in \hat{\J}$ \\ \hfill where $i \geq j$, and $| \Jh | = T$ \label{alg:majority:cons:j}
    \STATE $\hat{\J} \leftarrow \texttt{max\_indices}(\mathbf{z}, J)$
  \end{algorithmic}
  \mbox{\textit{Output:} $\Jh$ (observe that this is an estimate of $\T = \J$)}
\end{algorithm}

Studying Algorithm~\ref{alg:majority}, we see that the inputs are the support-set estimates from all sensors in the network, and the support-set cardinality. In the initialization phase, a large $N$-sized vector $\mathbf{z}$ is created; where the votes of the sensors are collected. Then, the estimate $\Jh$ is chosen based on the highest $T$ values in $\mathbf{z}$, which corresponds to majority voting. Observe that when knowledge is available about $J$, the \majority may be used also for the mixed support-set model, which we did (under another name) in \cite{Sundman:a_greedy_pursuit_algorithm}.

\subsection{Analysis}
When the nodes have found the support-set estimates by \gp algorithms, we use the input/output relation in Definition~\ref{ass} to provide some fundamental results valid for the \majority algorithm.

\begin{proposition} \label{prop:common_cons_l}
  In a setup with $L = h+m$ sensors with signal support-sets $\Thi_l$ for $l = 1,2, \dots, h$ and $\Thii_l$ for $l = 1,2, \dots, m$, let us assume that the index $i \in \left( \bigcap_{l = 1}^{h} \Thi_l, \bigcap_{l = 1}^{m} \Thii_l^{\complement} \right)$. Then, the \majority algorithm finds the estimate $\Jh$ such that $i \in \Jh$. In this case, the probability of detection is
  \begin{align}
    & \P{i\in\J | i\in \bigcap_{l = 1}^{h} \Thi_l,  i\in\bigcap_{l = 1}^{m} \Thii_l^{\complement}} \nonumber \\
    & \phantom{=} = \frac{(1-\epsilon)^h \epsilon^m J}{(1-\epsilon)^h \epsilon^m J + (\frac{T}{N-T}\epsilon)^h (1 - \frac{T}{N-T}\epsilon)^m (N-J)}, \label{eqn:common_cons_l}
  \end{align}
  where $J = T$.
  \begin{IEEEproof}
    Proof in Appendix~\ref{app:majority}.
  \end{IEEEproof}
\end{proposition}

\begin{figure}[t]
  \centering
  \includegraphics[width=\columnwidth]{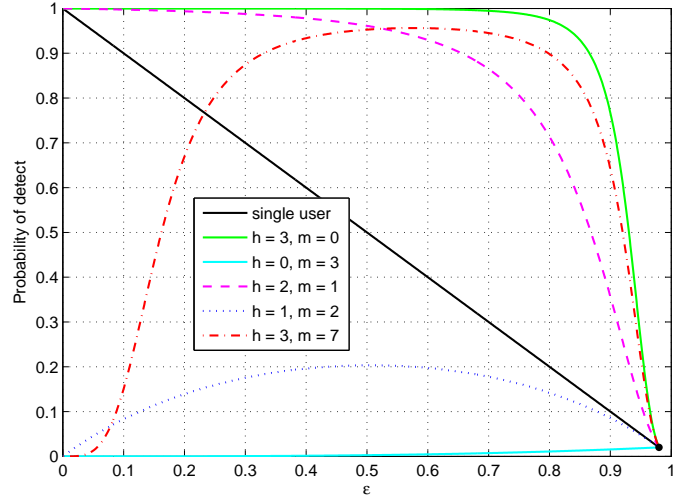}
  \caption{A few examples of probability of detect for the \majority algorithm using the common support-set model.} \label{fig:common_support}
\end{figure}

Getting any insight for the behavior of \majority from Proposition~\ref{prop:common_cons_l} is a non-trivial since \eqref{eqn:common_cons_l} is a complicated function of $m$, $h$, $J$, $T$ and $N$. For better understanding, we provide an example where some parameters are fixed.
\begin{example} \label{example1}
  Using $N = 1000$, $T = J = 20$, we provide \figurename~\ref{fig:common_support} where several pairs of $\{h,m\}$ are tested via \eqref{eqn:common_cons_l}. The black curve corresponds to the disconnected performance of Proposition~\eqref{prop:disconnect} and the black dot corresponds to the probability of detect at $\epsilon_{\max} = \frac{N-T}{N}$ which is the biggest value $\epsilon$ can take. Worth noticing in this figure is the interplay between hits and misses which may cause the performance to be very good at some parts, while being poor at other parts. This is illustrated in the curve for $h=3$, $m = 7$. An observation we found is that whenever $h>m$ we get good performance. \wqed
\end{example}

Using \majority voting, it is intuitively clear that more votes are always better (for a constant number of total sensors in the network). We show this explicitly with the following proposition.
\begin{proposition} \label{prop:common_cons_ineq} For the same setup as in Proposition~\ref{prop:common_cons_l} the following relation holds
  \begin{align}
  & \P{i\in\J | i\in\bigcap_{l = 1}^{h}\Thi_l, i\in\bigcap_{l = 1}^{m}\Thii_l^{\complement}} \nonumber \\
  & \hspace{2cm} \geq \P{i\in\J | i\in\bigcap_{l=1}^{h-1}\Thi_l, i\in\bigcap_{l=1}^{m+1}\Thii_l^{\complement}} \label{eqn:common_cons_ineq}
  \end{align}
  \begin{IEEEproof}
    Proof in Appendix~\ref{app:majority}.
  \end{IEEEproof}
\end{proposition}
By \propositionname~\ref{prop:common_cons_ineq}, it is clear that in a network of sensors, under the common support-set model, the \majority voting has a merit to detect the support-set.
%In this scenario, where a fusion center only has access to support-sets estimates from neighboring nodes, the only operation that can provide any insight in terms of reliability is counting overlaps in terms of voting (i.e., intersections). Therefore, it is an immediate implication of \propositionname~\ref{prop:common_cons_ineq} that a voting based strategy is in this case optimal. {\color{red} [Saikat: Should we make this claim; i.e., that voting is indeed optimal? I think it can be made. Maybe it then should be put in a Proposition. But it is not completely clear how to formulate such a proposition since the proof is somehow implicit from \propositionname~\ref{prop:common_cons_ineq} and the fact that there are no other possibilities than to count votes]}

\section{Mixed Support-set: \\ Distributed Parallel Pursuit} \label{sec:mixed_model}
We now consider the voting approach in a scenario where the signal correlation is modeled with the mixed support-set model. Assuming $T$ to be known (but not $J$), we previously developed such an algorithm in \cite{Sundman:dipp_arxiv}, where it is called \consensus voting. In this case, there is no fusion center; instead the sensors exchange support-set estimates and apply the \consensus algorithm locally, based on information from the neighboring sensors.

\subsection{Algorithm}
The \consensus algorithm differs from \majority since it has no knowledge of the support-set size of the joint support $J$. Instead it performs a threshold operation by selecting components with at least two votes. We have provided \consensus in Algorithm~\ref{alg:consensus}.
\begin{algorithm}[ht] 
\caption{\consensus: %\newline
\textit{Executed in the local node $\p$}} \label{alg:consensus} 
\textit{Input:} $\{ \hat{\T}_{\q} \}_{\q \in \mathcal{L}_{\p}^{\text{in}}}$, $\hat{\T}_{\p}$, $T$ \\
\textit{Initialization:} $\mathbf{z} \leftarrow \mathbf{0}_{N\times 1}$ \label{alg:cons:0} \\
\textit{Algorithm:}
\begin{algorithmic}[1]
\STATE $\mathbf{z} \leftarrow \addi(\mathbf{z}, \hat{\T}_{\p})$  \hspace{1cm} ($\p$-th node's estimate) \label{alg:cons:own}
\FOR { each $\q \in \mathcal{L}_{\p}^{\text{in}}$}
\STATE $\mathbf{z} \leftarrow \addi(\mathbf{z}, \hat{\T}_{\q})$ \hspace{1cm} (The neigbors' estimates) \label{alg:cons:others}
\ENDFOR
\STATE Choose $\hat{\J}_{\p}$ \hfill such that \hfill $(z(i) \geq 2) ~ \forall i \in \Jh_{\p}$ \\ \hfill and $| \Jh_{\p} | \leq T$ \label{alg:cons:j}
\end{algorithmic}
\mbox{\textit{Output:} $\Jh_{\p}$}
\end{algorithm}

Studying \algorithmname~\ref{alg:consensus}, the inputs are: a set of estimated support-sets $\{ \hat{\T}_{\q} \}_{\q \in \mathcal{L}^{\text{in}}}$ from the neighbors, the locally estimated support-set $\hat{\T}_{\p}$, and the sparsity level $T$. The estimate of $\hat{\J}_{\p}$ is formed (step~\ref{alg:cons:j}) such that no index in $\hat{\J}_{\p}$ has less than two votes (i.e., each index in $\hat{\J}_{\p}$ is present in at least two support-sets from $\{ \{ \hat{\T}_{\q} \}_{\q \in \mathcal{L}_{\p}^{\text{in}}}, \hat{\T}_{\p}\}$)\footnote{For node $\p$, this is equivalent to let algorithm choose $\Jh_{\p}$ as the union of all pair-wise intersections of support-sets (see the analysis section~\ref{sec:cons:analysis} for details).}. If the number of indices with at least two votes exceed the cardinality $T$, we pick the $T$ largest indices.  Thus, the \consensus strategy can be summarized as:
\begin{itemize}
\item Pick indices for $\Jh_{\p}$ that have two votes
\item If $|\Jh_{\p}| > T$, choose the $T$ largest indices
\end{itemize}
In the following we will analyze the \consensus strategy using the input/output relation of Definition~\ref{ass}.

\subsection{Analysis} \label{sec:cons:analysis}
Assuming the nodes use \gp algorithms to find the support-set estimates, we obtain the following results.
\begin{proposition} \label{cons:prop:pi}
  The probability that an index `$i$' is correct for sensor `$\p$', provided that this index is detected by the sensor `$\p$' itself and additionally `$h$' neighbors, but not detected by `$m$' neighbors is given by \eqref{eqn:pi_phi_phii_phiiic}.
  \begin{IEEEproof}
    Proof in Appendix~\ref{app:consensus}.
  \end{IEEEproof}
\end{proposition}

\begin{proposition} \label{cons:prop:pic}
  The probability that an index `$i$' is correct for sensor `$\p$', provided that this index is detected by `$h$' neighbors, but not detected by the sensor `$\p$' itself and additionally `$m$' neighbors is given by \eqref{eqn:pi_phic_phii_phiiic}.
  \begin{IEEEproof}
    Proof in Appendix~\ref{app:consensus}.
  \end{IEEEproof}
\end{proposition}

\begin{figure*}[!t]
\normalsize
\begin{align}
  & \P{i\in\Ti | i\in\Thi, i\in\bigcap_{l = 1}^{h}\Thii_l, i\in\bigcap_{l = 1}^{m}\Thiii_l^{\complement}} =  \label{eqn:pi_phi_phii_phiiic} \\
  & \frac{(1\!-\!\epsilon)^{h\!+\!1} \epsilon^{m} \frac{J}{N} \!+\! (1\!-\!\epsilon) (\frac{T}{N\!-\!T}\epsilon)^{h} (1 \!-\! \frac{T}{N\!-\!T}\epsilon)^{m} \frac{I}{N} }
  {(1\!-\!\epsilon)^{h\!+\!1} \epsilon^{m} \frac{J}{N} \!+\! (h+1) (1\!-\!\epsilon) (\frac{T}{N\!-\!T}\epsilon)^{h}(1\!-\!\frac{T}{N\!-\!T}\epsilon)^m \frac{I}{N} \!+\! m \epsilon (\frac{T}{N\!-\!T}\epsilon)^{h\!+\!1} (1\!-\!\frac{T}{N\!-\!T}\epsilon)^{m\!-\!1} \frac{I}{N} \!+\! (\frac{T}{N\!-\!T}\epsilon)^{h\!+\!1} (1\!-\!\frac{T}{N\!-\!T}\epsilon)^{m} \frac{N\!-\!J\!-\!(m\!+\!h\!+\!1)I}{N}}. \nonumber
\end{align}

\begin{align}
  & \P{i\in\Ti | i\in\Thi^{\complement}, i\in\bigcap_{l = 1}^{h}\Thii_l, i\in\bigcap_{l = 1}^{m}\Thiii_l^{\complement}} =  \label{eqn:pi_phic_phii_phiiic} \\
  & \frac{(1\!-\!\epsilon)^h\epsilon^{m\!+\!1}\frac{J}{N} \!+\! \epsilon (\frac{T}{N\!-\!T}\epsilon)^h (1\!-\!\frac{T}{N\!-\!T}\epsilon)^m\frac{I}{N}}{(1\!-\!\epsilon)^{h} \epsilon^{m\!+\!1} \frac{J}{N} \!+\! h (1\!-\!\epsilon) (\frac{T}{N\!-\!T}\epsilon)^{h\!-\!1}(1\!-\!\frac{T}{N\!-\!T}\epsilon)^{m\!+\!1} \frac{I}{N} \!+\! (m\!+\!1) \epsilon (\frac{T}{N\!-\!T}\epsilon)^{h} (1\!-\!\frac{T}{N\!-\!T}\epsilon)^{m} \frac{I}{N} \!+\! (\frac{T}{N\!-\!T}\epsilon)^{h} (1\!-\!\frac{T}{N\!-\!T}\epsilon)^{m\!+\!1} \frac{N \!-\! J \!-\!(m\!+\!h\!+\!1)I}{N}}. \nonumber
\end{align}
% IEEE uses as a separator
\hrulefill
% The spacer can be tweaked to stop underfull vboxes.
\vspace*{4pt}
\end{figure*}

\begin{figure}[t]
  \centering
  \includegraphics[width=\columnwidth]{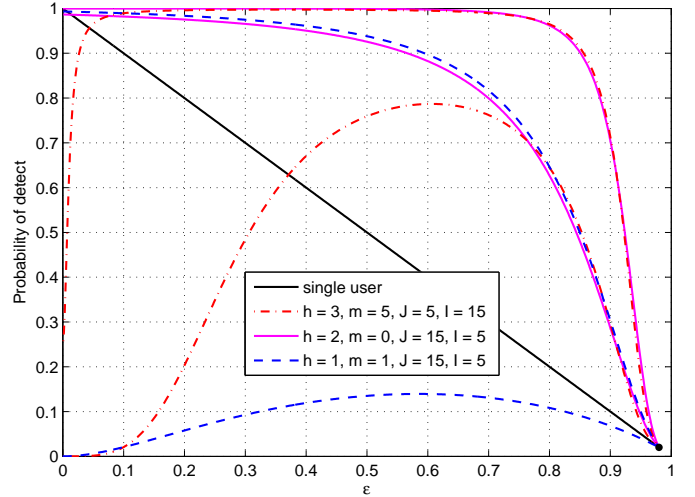}
  \caption{Analytical results for the mixed support-set model. Observe here that there is always a total number of $h+m+1$ nodes present for a \consensus algorithm.} \label{fig:mixed_support}
\end{figure}

Getting any insight from these propositions is difficult. Therefore, we provide the following numerical example.
\begin{example} \label{example2}
  In \figurename~\ref{fig:mixed_support} we provide examples for the mixed support-set model using Proposition~\ref{cons:prop:pi} and Proposition~\ref{cons:prop:pic}. In this system $N = 1000$ and $\epsilon$ is varied.
Notice in \figurename~\ref{fig:mixed_support}, that there are two curves for each configuration. The top-most curve corresponds to \eqref{eqn:pi_phi_phii_phiiic}, where the sensor `$\p$' itself found the index, and the lower-most curve corresponds to \eqref{eqn:pi_phic_phii_phiiic}, where the sensor `$\p$' itself missed the index.
%Notice that the curves come in pairs, where the upper curve corresponds to \eqref{eqn:pi_phi_phii_phiiic} and is based on that $h$ neighbors and local node index detection, while the lower curve corresponds to \eqref{eqn:pi_phic_phii_phiiic} and is based on that $h$ neighbors but the local node missed the index.
By testing it seems that, similarly to \majority voting, when $h > m$, the performance is good.
\end{example}

Derivation of further general results based on Proposition~\ref{cons:prop:pi} and Proposition~\ref{cons:prop:pic}, for example providing general precise requirements under which \consensus provides higher probability than a single sensor case is non-trivial. Instead, we assume a limited number of neighbors and fix a number of parameters according to \cite{Sundman:dipp_arxiv}. In particular, we assume that each local node $\p$ has two independent neighbors. Using information obtained by the neighbors and the local node, the \consensus endeavors to estimate the joint support part as $\Jh_{\p}$. Following \algorithmname~\ref{alg:consensus}, we note that in this case
\begin{align}
  \Jh_{\p} = \Bigl\{ i: i\in & \left( (\Thi \cap \Thii \cap \Thiii) \cup (\Thi \cap \Thii \cap \Thiii^{\complement}) \right. \nonumber \\
      & \left. \phantom{=} \cup (\Thi^{\complement} \cap \Thii \cap \Thiii) \right) \,\, \forall \q,\r \in \mathcal{L}_{\p}^{\text{in}}, \q \neq \r \Bigr\}. \label{eqn:jh}
\end{align}

Using \eqref{eqn:jh}, we provide the following remark in numerical manner.

\begin{remark} \label{prop:ineq_example}
  When $N = 1000$, $T = 20$, $J = 15$, $V = 2$ and $0.0140 \leq \epsilon \leq \epsilon_{\max} = 0.98$, then the probability of $i\in\Jh_{\p}$ to be correct is always bigger than or equal to the probability of $i\in\Thi$ to be correct, that is
  \begin{align}
    \P{i\in\Ti | i\in\Jh_{\p}} \geq \P{i\in\Ti | i\in\Thi}.
  \end{align}

  \begin{IEEEproof}
    Proof in Appendix~\ref{app:consensus}.
  \end{IEEEproof}
\end{remark}

 We note that although Remark~\ref{prop:ineq_example} strongly suggest that the \majority voting provides for a good result, we can consider the typical \cs condition that $N$ is very large. Then an even stronger result can be formulated as in the following corollary.
\begin{corollary} \label{cons:coro:inequalities}
  If $J \geq 1$, $T$ grows sublinearly in $N$, and $\Jh_{\p}$ is the output of \consensus, then
  \begin{align}
    \lim_{N \rightarrow \infty} \P{i\in\Ti|i\in\Jh_{\p}} = 1.
  \end{align}
  \begin{IEEEproof}[Proof of Corollary~\ref{cons:coro:inequalities}]
    For this proof, we show that \eqref{eqn:pi_phi_phii_phiiic} tends to one when $h \geq 1$ (follows from $\Jh_{\p}$) and that \eqref{eqn:pi_phic_phii_phiiic} tends to one when $h \geq 2$ (also follows from $\Jh_{\p}$).

    First consider \eqref{eqn:pi_phi_phii_phiiic} and note that it can never happen that $m > 0$ when $\epsilon = 0$. Then it is straight-forward to see that $(\frac{T}{N-T}\epsilon)^h \rightarrow 0$ since $h \geq 1$, hence the whole expression tends to 1.

    Similarly for \eqref{eqn:pi_phic_phii_phiiic}, note that it can never happen that $m > 0$ when $\epsilon = 0$. Then it is straight-forward to see that $(\frac{T}{N-T}\epsilon)^{h-1} \rightarrow 0$ since $h \geq 2$, consequently the whole expression tends to 1.
  \end{IEEEproof}
\end{corollary}

\section{Experimental Evaluation} \label{sec:experiment}
In this section we perform two experiments to illustrate the three results: Proposition~\ref{prop:common_cons_l}, Proposition~\ref{cons:prop:pi} and Proposition~\ref{cons:prop:pic}. The goal of these experiments is to compare the analytical results with observations from a simulation process. Since there is no closed form result for $\epsilon$; this entity has to be estimated. We estimate $\epsilon$ in the same way as in the second test of Section~\ref{sec:evaluation_prob_model}, and by averaging over all signals. To find the performance of the different voting strategies, we count how many times `$h$' hits and `$m$' misses correspond to a correct support-set index estimate and divide this number with the number of times `$h$' hits and `$m$' misses occurs in total. Thus the procedure is as follows:
\begin{enumerate}
\item Estimate $\hat{\epsilon}$.
\item For each $\hat{\epsilon}$, count the actual accuracy of the voting procedure and put a mark at this point.
\item Compare to the theoretical expression in the respective equation.
\end{enumerate}

In \figurename~\ref{fig:common_support}, $\epsilon$ is plotted vs the probability of detection for the results of the common support-set model. A total number of $10^6$ random simulations are performed, using parameters $N = 1000$, $T = 20$ and \gp algorithm subspace pursuit (\sp). To find different $\epsilon$, $M$ and $\smnr$ are varied. In the case where $h = 2, m = 1$, the $\epsilon$ from left to right are found by: $M = 96, 85, 76, 64, 50, 41, 34, 28$ with corresponding $\smnr = 20, 20, 20, 10, 10, 10, 10, 0$ and for the case where $h = 3, m = 7$, $M = 101, 96, 92, 88, 50, 41, 34, 28$ with corresponding $\smnr = 20, 20, 20, 20, 10, 10, 10, 0$. Observe that largest possible $\epsilon_{\max} = \frac{N-T}{N}$ (marked with a small black dot). The equations used for the analytical results are found in \eqref{eqn:common_cons_l}. When we compare the simulations to the result predicted by analysis, we notice an almost perfect match. We argue that the slight mismatch for some points is due to noise and will average out using a larger simulation ensemble. For example, it is a rare event that $h = 3$ $m= 7$ occurs when the algorithms are very good (i.e., the simulation point at $(0.05, 0.05)$).
\begin{figure}[t]
  \centering
  \psfrag{aaaaaaaaaaaaaaaaaa}{\scriptsize $\sP{i \in \J | i \in \Thi}$}
  \psfrag{bbbbbbbbbbbbbbbbbb}{\scriptsize \textsf{simulation}}
  \psfrag{cccccccccccccccccc}{\scriptsize $\sP{i \in \J | i \in \cup_{\p = 1}^2 \Thi}$}
  \psfrag{dddddddddddddddddd}{\scriptsize \textsf{simulation}}
  \psfrag{eeeeeeeeeeeeeeeeee}{\scriptsize $\sP{i \in \J | i \in \cup_{\p = 1}^4 \Thi}$}
  \psfrag{ffffffffffffffffff}{\scriptsize \textsf{simulation}}
  \includegraphics[width=\columnwidth]{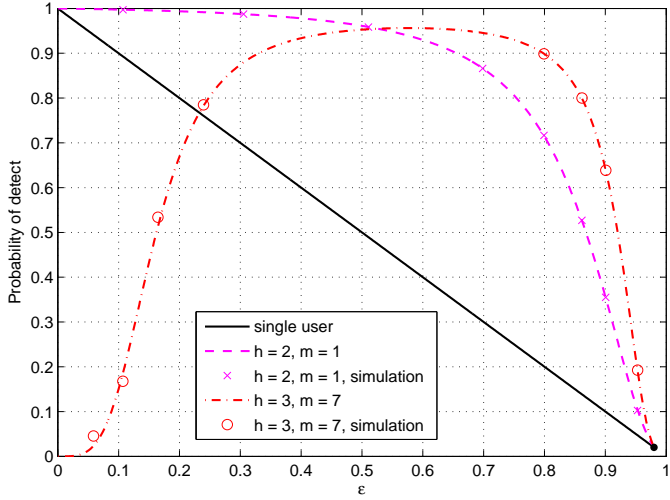}
  \caption{Analytical and simulation results for the voting performance based on the common support-set model.} \label{fig:common_support_simulation}
\end{figure}

In \figurename~\ref{fig:mixed_support}, $\epsilon$ is plotted vs the probability of detection for the mixed support-set model. A total number of $10^6$ random simulations are performed, using parameters $N = 1000$ and \gp algorithm subspace pursuit (\sp). To find different $\epsilon$, $M$ and $\smnr$ are varied. Here we used the same data-points for all curves; the $\epsilon$ from left to right: $M = 96, 90, 85, 76, 64, 50, 41, 34, 28$ with corresponding $\smnr = 20, 20, 20, 20, 10, 10, 10, 10, 0$. We observe that also here, the simulation points match closely to the predicted values. The equations used for the analytical results are found in \eqref{eqn:pi_phi_phii_phiiic} and \eqref{eqn:pi_phic_phii_phiiic}.
\begin{figure}[t]
  \centering
  \psfrag{aaaaaaaaaaaaaaaaaaaa}{\scriptsize $\sP{i \in \Ti | i \in \Thi}$}
  \psfrag{bbbbbbbbbbbbbbbbbbbb}{\scriptsize \textsf{simulation}}
  \psfrag{cccccccccccccccccccc}{\scriptsize $\sP{i \in \Ti | i \in \Thii, i\in\Thiii}$}
  \psfrag{dddddddddddddddddddd}{\scriptsize \textsf{simulation}}
  \psfrag{eeeeeeeeeeeeeeeeeeee}{\scriptsize $\sP{i \in \Ti | i \in \Thi, i\in\Thii}$}
  \psfrag{ffffffffffffffffffff}{\scriptsize \textsf{simulation}}
  \includegraphics[width=\columnwidth]{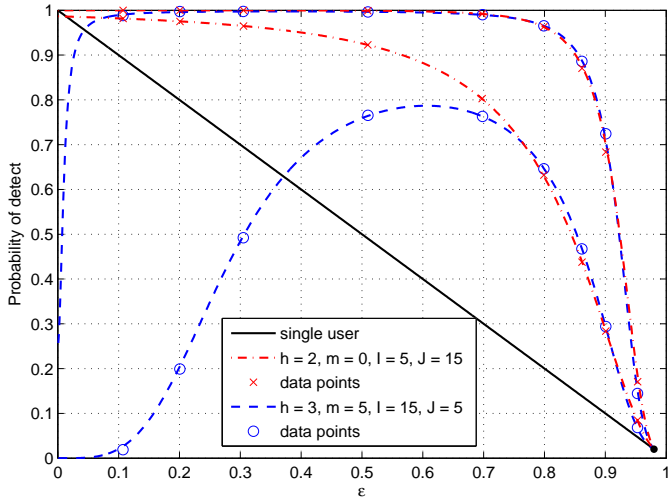}
  \caption{Analytical and simulation results for the \consensus performance based on the mixed support-set model.} \label{fig:mixed_support_simulation}
\end{figure}

\section{Conclusion}
In this paper, we have analyzed democratic based voting strategies for support-sets estimation using greedy algorithms. We have characterized the input/output relation of any typical \gp algorithm based on four relations. Using these relations we shown the merit of voting for two particular examples: the \majority algorithm and the \consensus algorithm, both which has been presented in the literature earlier. With several experiments, we validated both the input/output relation and the results derived; in all cases the experiments closely matched the theoretical prediction.

\appendices
\section{Proofs}
Here, we present proofs for the propositions provided in the paper. First, we introduce some lemmas used in the proofs. Then, in \appendixname~\ref{app:majority} we present the proofs for \propositionname~\ref{prop:common_cons_l} and \propositionname~\ref{prop:common_cons_ineq}; in \appendixname~\ref{app:consensus} we present the proofs for \propositionname~\ref{cons:prop:pi}, \propositionname~\ref{cons:prop:pic} and remark~\ref{prop:ineq_example}.

\begin{lemma}[Equi-probability of Subsets] \label{lemma:inverted}
  For any $\mathcal{A} \subseteq \Ti$ and for any $\mathcal{B} \subseteq \Ti^{\complement}$, the following holds
  \begin{align}
    \P{ i\in\mathcal{A} | i\in\Thi } & 
    = \frac{|\mathcal{A}|}{T}\P{ i\in\Thi | i\in\Ti },
    \label{lemma:inverted_true1}  \\ %cons:prop1
    \P{ i\in\Thi | i\in\mathcal{A} } &
    = \P{ i\in\Thi | i\in\Ti },
    \label{lemma:inverted_true2} \\ %cons:prop4
    \P{ i\in\mathcal{B} | i\in\Thi } & 
    = \frac{|\mathcal{B}|}{T}\P{ i\in\Thi | i\in\Ti^{\complement} },
    \label{lemma:inverted_false1} \\ %\label{cons:prop5}
    \P{ i\in\Thi | i\in\mathcal{B} } &
    = \P{ i\in\Thi | i\in\Ti^{\complement} }.
    \label{lemma:inverted_false2} %\label{cons:prop5}
  \end{align}
  \begin{IEEEproof}[Proof for \eqref{lemma:inverted_true1}]
    \begin{align}
      \P{ i\in\mathcal{A} | i\in\Thi } &
      \overset{(a)}{=} \frac{|\mathcal{A}|}{T} \P{i \in \Ti | i \in \Thi} \nonumber \\
      & = \frac{|\mathcal{A}|}{T} \P{i \in \Thi | i \in \Ti} \frac{\P{i \in \Ti}}{\P{i\in \Thi}} \nonumber \\
      & = \frac{|\mathcal{A}|}{T} \P{i \in \Thi | i \in \Ti} \nonumber
    \end{align}
    Here, \ensuremath{(a)} follows directly from Definition~\ref{ass}.
  \end{IEEEproof}

  \begin{IEEEproof}[Proof for \eqref{lemma:inverted_true2}]
    \begin{align}
      \P{ i\in\Thi | i\in\mathcal{A} } & = \P{ i\in\mathcal{A} | i\in\Thi } \frac{ \P{i\in\Thi} }{ \P{i\in\mathcal{A} } } \nonumber \\
      & \overset{(a)}{=} \frac{|\mathcal{A}|}{T}\P{ i\in\Thi | i\in\Ti }\frac{ \P{i\in\Thi} }{ \P{i\in\mathcal{A} } } \nonumber  \\
      & = \P{ i\in\Thi | i\in\Ti }, \nonumber
    \end{align}
    where we in \ensuremath{(a)} applied \eqref{lemma:inverted_true1}.
  \end{IEEEproof}

  \begin{IEEEproof}[Proof for \eqref{lemma:inverted_false1}] This proof is similar to the proof for \eqref{lemma:inverted_true1},
    \begin{align}
      \P{ i\in\mathcal{B} | i\in\Thi }  
      & = \frac{|\mathcal{B}|}{|\Ti^{\complement}|}\P{i\in\Ti^{\complement}|i\in\Thi} \nonumber \\
      & = \frac{|\mathcal{B}|}{|\Ti^{\complement}|}\P{i\in\Thi|i\in\Ti^{\complement}}\frac{|\Ti^{\complement}|}{|\Thi|} \nonumber \\
      & = \frac{|\mathcal{B}|}{T}\P{ i\in\Thi | i\in\Ti^{\complement} }. \nonumber
    \end{align}
%    Notice that this expression actually is $\frac{|\mathcal{B}|}{T}\P{ i\in\Thi | i\in\Ti^{\complement} } = \epsilon$ according to Definition~\ref{ass}.
  \end{IEEEproof}

  \begin{IEEEproof}[Proof for \eqref{lemma:inverted_false2}]
    This proof is similar to the proof for \eqref{lemma:inverted_true2} and follows directly by applying \eqref{lemma:inverted_false1}.% ,
    % \begin{align}
    %   \P{ i\in\Thi | i\in\mathcal{B} }
    %   & = \P{ i\in\mathcal{B} | i\in\Thi }\frac{|\Thi|}{|\mathcal{B}|} \nonumber \\
    %   & \overset{(a)}{=} \P{ i\in\Thi | i\in\Ti^{\complement} }, \nonumber
    % \end{align}
    % where \ensuremath{(a)} follows from \eqref{lemma:inverted_false1}.
  \end{IEEEproof}
\end{lemma}

\begin{lemma}[Independence of Joint Probability] \label{lemma:indep}
  The local results from different sensor nodes are independent over certain regions. Assume there are a total of $h+m$ nodes in the system and that we denote different nodes by sub-indices $\p_k \neq \q_l ~ \forall k,l$ and $\p \neq \p_k, \p\neq \q_l ~ \forall k,l $. Then, for $\mathcal{A} \subseteq \J$, $\mathcal{B} \subseteq \Ii_h$, $\mathcal{C} \subseteq \Iii_m$, and $\mathcal{D} \subseteq (\J \cup \bigcup_{l=1}^h \Ii_l \cup \bigcup_{l=1}^m \Iii_l)^{\complement}$ the following relations hold:
  \begin{align}
    & \P{i\in\bigcap_{l=1}^{h} \Thi_l, i\in\bigcap_{l=1}^{m} \Thii_l^{\complement} | i\in\mathcal{A}} \label{lemma:indep1} \\
    & = \P{i\in\Thi| i\in\Ti}^{h} \P{i\in \Thi^{\complement} | i\in\Ti}^m, \nonumber \\
    & \P{i\in\bigcap_{l=1}^{h} \Thi_l, i\in\bigcap_{l=1}^{m} \Thii_l^{\complement} | i\in\mathcal{B}} \label{lemma:indep2}\\
    & = \P{i\in\Thi| i\in\Ti}\P{i\in\Thi| i\in\Ti^{\complement}}^{h-1} \P{i\in \Thi^{\complement} | i\in\Ti^{\complement}}^m, \nonumber \\
    & \P{i\in\bigcap_{l=1}^{h} \Thi_l, i\in\bigcap_{l=1}^{m} \Thii_l^{\complement} | i\in\mathcal{C}} \label{lemma:indep3} \\
    & = \P{i\in\Thi^{\complement}| i\in\Ti}\P{i\in\Thi| i\in\Ti^{\complement}}^{h} \P{i\in \Thi^{\complement} | i\in\Ti^{\complement}}^{m-1}, \nonumber \\
    & \P{i\in\bigcap_{l=1}^{h} \Thi_l, i\in\bigcap_{l=1}^{m} \Thii_l^{\complement} | i\in\mathcal{D}} \label{lemma:indep4} \\
    & = \P{i\in\Thi| i\in\Ti^{\complement}}^{h} \P{i\in \Thi^{\complement} | i\in\Ti^{\complement}}^m. \nonumber
  \end{align}

  \begin{IEEEproof}[Proof of \eqref{lemma:indep1}]
    \begin{align}
      & \P{i\in\bigcap_{l=1}^{h} \Thi_l, i\in\bigcap_{l=1}^{m} \Thii_l^{\complement} | i\in\mathcal{A}} \nonumber \\
%      & \overset{(a)}{=} \P{i\in\bigcap_{l=1}^{h} \Thi_l |  i\in\bigcap_{l=1}^{m} \Thii_l^{\complement}, i\in\mathcal{A}} \P{i\in\bigcap_{l=1}^{m} \Thii_l^{\complement} | i\in\mathcal{A}} \nonumber \\
      & \overset{(a)}{=} \prod_{l=1}^{h} \P{i\in\Thi_l| i\in\mathcal{A}} \prod_{l=1}^m \P{i\in \Thii_l^{\complement} | i\in\mathcal{A}}, \nonumber \\
      & \overset{(b)}{=} \prod_{l=1}^{h} \P{i\in\Thi_l| i\in\Ti_l} \prod_{l=1}^m \P{i\in \Thii_l^{\complement} | i\in\Tii_l}, \nonumber \\
      & \overset{(c)}{=} \P{i\in\Thi| i\in\Ti}^{h} \P{i\in \Thi^{\complement} | i\in\Ti}^m, \nonumber
    \end{align}
    In \ensuremath{(a)} we applied the chain rule on all intersections and applying the Markov property, which is illustrated in \figurename~\ref{fig:app:flow}. In \ensuremath{(b)} we have used Lemma~\ref{lemma:inverted} and in \ensuremath{(c)} we have used that all probabilities are equal.    
  \end{IEEEproof}
  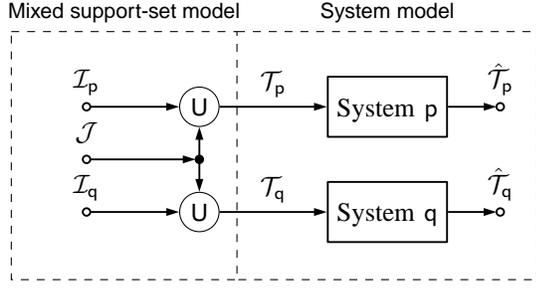
\begin{figure}[t]
    \centering
    \begin{tikzpicture}[node distance=1.5cm]
      \node[dspnodeopen,dsp/label=above] (Ip) {$\Ii$};
      \node[dspnodeopen,dsp/label=above, below of=Ip, node distance=0.7cm] (J) {$\J$};
      \node[dspnodeopen,dsp/label=above, below of=J, node distance=0.7cm] (Iq) {$\Iii$};
      \node[draw,circle,right of=Ip,inner sep=2pt] (add1) {\small\textsf{U}};
      \node[dspnodefull, right of=J] (divider) {};
      \node[draw,circle,right of=Iq,inner sep=2pt] (add2) {\small\textsf{U}};
      \node[dspsquare, right of=add1, node distance=2.5cm] (s1) {~System~$\p$~};
      \node[dspsquare, right of=add2, node distance=2.5cm] (s2) {~System~$\q$~};
      \node[dspnodeopen,dsp/label=above, right of=s1] (Tp) {$\Thi$};
      \node[dspnodeopen,dsp/label=above, right of=s2] (Tq) {$\Thii$};
      \draw[dspconn] (Ip) -- (add1);
      \draw[dspconn] (Iq) -- (add2);
      \draw[dspconn] (J) -- (divider);
      \draw[dspconn] (divider) -- (add1);
      \draw[dspconn] (divider) -- (add2);
      \draw[dspconn] (add1) -- node[midway,above] {$\Ti$} (s1);
      \draw[dspconn] (add2) -- node[midway,above] {$\Tii$} (s2);
      \draw[dspconn] (s1) -- (Tp);
      \draw[dspconn] (s2) -- (Tq);
      \draw[dashed] (-1cm,1cm) -- node[midway,above] {\sffamily\footnotesize Mixed support-set model} (2cm, 1cm) -- (2cm,-2.3cm)  -- (-1cm, -2.3cm) -- (-1cm,1cm);
      \draw[dashed] (2cm,1cm) -- node[midway,above] {\sffamily\footnotesize\phantom{p}System model\phantom{p}} (6cm,1cm) -- (6cm,-2.3cm) -- (2cm,-2.3cm);
    \end{tikzpicture}
    \caption{For two nodes, this figure illustrates the first order Markov property of the outputs $\hat{\T}_{\p}$ and $\hat{\T}_{\q}$.}
    \label{fig:app:flow}
  \end{figure}
  \begin{IEEEproof}[Proof of \eqref{lemma:indep2}]
    \begin{align}
      & \P{i\in\bigcap_{l=1}^{h} \Thi_l, i\in\bigcap_{l=1}^{m} \Thii_l^{\complement} | i\in\mathcal{B}} \\
      & = \P{i\in\Thi_h, i\in\bigcap_{l=1}^{h-1} \Thi_l, i\in\bigcap_{l=1}^{m} \Thii_l^{\complement} | i\in\mathcal{B}} \nonumber \\
      & = \P{i\in\Thi_h | i\in\mathcal{B}} \prod_{l=1}^{h-1} \P{i\in \Thi_l| i\in\mathcal{B}} \nonumber \\
      & \phantom{=} \cdot \prod_{l=1}^{m} \P{i\in\bigcap \Thii_l^{\complement} | i\in\mathcal{B}} \nonumber \\
      & \overset{(c)}{=} \P{i\in\Thi_h | i\in\Ti_h} \prod_{l=1}^{h-1} \P{i\in \Thi_l| i\in\Ti_l^{\complement}} \nonumber \\
      & \phantom{=} \cdot \prod_{l=1}^{m}\P{i\in\bigcap \Thii_l^{\complement} | i\in\Tii_l^{\complement}} \nonumber \\
      & \overset{(d)}{=} \P{i\in\Thi| i\in\Ti}\P{i\in\Thi| i\in\Ti^{\complement}}^{h-1} \P{i\in \Thi^{\complement} | i\in\Ti^{\complement}}^m. \nonumber
    \end{align}
    In \ensuremath{(a)}, Lemma~\ref{lemma:inverted} is used, and in \ensuremath{(b)} we used that all probabilities are equal.
  \end{IEEEproof}
  
  \begin{IEEEproof}[Proof of \eqref{lemma:indep3}]
    The proofs is similar to the proof for \eqref{lemma:indep2}.
  \end{IEEEproof}

  \begin{IEEEproof}[Proof of \eqref{lemma:indep4}]
    The proofs is similar to the proof for \eqref{lemma:indep1}.
  \end{IEEEproof}
\end{lemma}

\begin{lemma}[Joint Probability]\label{lemma:joint_prob}
  Assume there are $h+m$ nodes in the system, that $\p_k \neq \q_l ~ \forall k,l$ are different nodes. Then, the following holds:
  \begin{align}
    & \P{i\in\bigcap_{l = 1}^{h}\Thi_l, i\in\bigcap_{l = 1}^{m}\Thii_l^{\complement}} \nonumber \\
    & = (1-\epsilon)^{h} \epsilon^{m} \frac{J}{N} \nonumber \\
    & \phantom{=} + h (1-\epsilon) (\frac{T}{N-T}\epsilon)^{h-1}(1-\frac{T}{N-T}\epsilon)^m \frac{I}{N} \nonumber \\
    & \phantom{=} + m \epsilon (\frac{T}{N-T}\epsilon)^{h} (1-\frac{T}{N-T}\epsilon)^{m-1} \frac{I}{N} \nonumber \\
    & \phantom{=} + (\frac{T}{N-T}\epsilon)^{h} (1-\frac{T}{N-T}\epsilon)^{m} \frac{N - J -(m+h)I}{N} \nonumber
  \end{align}
  \begin{IEEEproof}
    For this proof, we first introduce a notational simplification, define $\mathcal{U} = \J \cup \Ii_1 \cup \dots \cup \Ii_h \cup \Iii_1 \cup \dots \cup \Iii_m$. Observe that the sub-sets in $\mathcal{U}$ are non-overlapping. Then,
\begin{align}
  & \P{i\in\bigcap_{l = 1}^{h}\Thi_l, i\in\bigcap_{l = 1}^{m}\Thii_l^{\complement}} \nonumber \\
  & \overset{(a)}{=} \sum_{ \hspace{-8pt}\mathcal{A} = \substack{\J, \\
      \Ii_1, \Ii_2, \dots, \Ii_{h}, \\ 
      \Iii_1, \Iii_2, \dots, \Iii_{m}, \\
      \mathcal{U}^{\complement}}} \hspace{-20pt} \P{i\in\bigcap_{l = 1}^{h}\Thi_l, i\in\bigcap_{l = 1}^{m}\Thii_l^{\complement}| i\in\mathcal{A}} \P{i\in\mathcal{A}} \nonumber \\
  & \overset{(b)}{=} \P{i\in\bigcap_{l = 1}^{h}\Thi_l, i\in\bigcap_{l = 1}^{m}\Thii_l^{\complement}| i\in\J} \P{i\in\J} \nonumber \\
  & \phantom{=} + h \P{i\in\Thi,i\in\bigcap_{l = 1}^{h-1}\Thii_l, i\in\bigcap_{l = 1}^{m}\Thiii_l^{\complement}| i\in\Ii} \P{i\in\Ii} \nonumber \\
  & \phantom{=} + m \P{i\in\Thi^{\complement},i\in\bigcap_{l = 1}^{h}\Thii_l, i\in\bigcap_{l = 1}^{m-1}\Thiii_l^{\complement}| i\in\Ii} \P{i\in\Ii} \nonumber \\
  & \phantom{=} + \P{i\in\bigcap_{l = 1}^{h}\Thi_l, i\in\bigcap_{l = 1}^{m}\Thii_l^{\complement}| i\in\mathcal{U}^{\complement}} \P{i\in\mathcal{U}^{\complement}} \nonumber \\
  % & = \P{i\in\Thi | i\in\J}^{h} \P{i\in\Thii^{\complement}| i\in\J}^{m} \P{i\in\J} \nonumber \\
  % & \phantom{=} + h \P{i\in\Thi | i\in\Ii} \P{i\in\Thii|i\in\Ii}^{h-1} \nonumber \\
  % & \phantom{=} \phantom{=} \cdot \P{i\in\Thiii^{\complement}| i\in\Ii}^m \P{i\in\Ii} \nonumber \\
  % & \phantom{=} + m \P{i\in\Thi^{\complement}| i\in\Ii} \P{i\in\Thii| i\in\Ii}^{h} \nonumber \\
  % & \phantom{=} \phantom{=} \cdot \P{i\in\Thiii^{\complement}| i\in\Ii}^{m-1} \P{i\in\Ii} \nonumber \\
  % & \phantom{=} + \P{i\in\Thi| i\in\mathcal{U}^{\complement}}^{h} \P{i\in\Thii^{\complement}| i\in\mathcal{U}^{\complement}}^{m} \P{i\in\mathcal{U}^{\complement}} \nonumber \\
  & \overset{(c)}{=} \P{i\in\Thi | i\in\Ti}^{h} \P{i\in\Thii^{\complement}| i\in\Tii}^{m} \P{i\in\J} \nonumber \\
  & \phantom{=} + h \P{i\in\Thi | i\in\Ti} \P{i\in\Thii|i\in\Tii^{\complement}}^{h-1} \nonumber \\
  & \phantom{=} \phantom{=} \cdot \P{i\in\Thiii^{\complement}| i\in\Tiii^{\complement}}^m \P{i\in\Ii} \nonumber \\
  & \phantom{=} + m \P{i\in\Thi^{\complement}| i\in\Ti} \P{i\in\Thii| i\in\Tii^{\complement}}^{h} \nonumber \\
  & \phantom{=} \phantom{=} \cdot \P{i\in\Thiii^{\complement}| i\in\Tiii^{\complement}}^{m-1} \P{i\in\Ii} \nonumber \\
  & \phantom{=} + \P{i\in\Thi| i\in\Ti^{\complement}}^{h} \P{i\in\Thii^{\complement}| i\in\Tii^{\complement}}^{m} \P{i\in\mathcal{U}^{\complement}} \nonumber \\
  & \overset{(d)}{=} (1-\epsilon)^{h} \epsilon^{m} \frac{J}{N} \nonumber \\
  & \phantom{=} + h (1-\epsilon) (\frac{T}{N-T}\epsilon)^{h-1}(1-\frac{T}{N-T}\epsilon)^m \frac{I}{N} \nonumber \\
  & \phantom{=} + m \epsilon (\frac{T}{N-T}\epsilon)^{h} (1-\frac{T}{N-T}\epsilon)^{m-1} \frac{I}{N} \nonumber \\
  & \phantom{=} + (\frac{T}{N-T}\epsilon)^{h} (1-\frac{T}{N-T}\epsilon)^{m} \frac{N - J -(m+h)I}{N}. \nonumber
\end{align}
In \ensuremath{(a)}, the probability is marginalized over all individual and joint support-sets, and over $\mathcal{U}$. In \ensuremath{(b)}, we extend the sum. In \ensuremath{(c)} we apply Lemma~\ref{lemma:indep}. Lastly, in \ensuremath{(d)} we plug the values from Definition~\ref{ass}.
\end{IEEEproof}
\end{lemma}

\subsection{Proofs of the Results for \majority} \label{app:majority}
Here, we prove Proposition~\ref{prop:common_cons_l} and Proposition~\ref{prop:common_cons_ineq}, which are stated based on the common support-set model. Recall that in the common support-set model, $\J = \T$ and $J = T$. %in the \majority voting for the common support-set model.

\begin{IEEEproof}[Proof of Proposition~\ref{prop:common_cons_l}]
  \begin{align}
    & \P{i\in\J | i\in \bigcap_{l = 1}^{h} \Thi_l,  i\in\bigcap_{l = 1}^{m} \Thii_l^{\complement}} \nonumber \\
    & = \frac{\P{i\in\bigcap_{l=1}^{h} \Thi_l, i\in\bigcap_{l = 1}^{m} \Thii_l^{\complement} | i\in\J } \P{i\in\J}}{\P{i\in\bigcap_{l = 1}^{h} \Thi_l, i\in\bigcap_{l =1}^{m} \Thii_l^{\complement}}} \nonumber \\
    & = \frac{\P{i\in\bigcap_{l = 1}^{h} \Thi_l, i\in\bigcap_{l=1}^{m} \Thii_l^{\complement} | i\in\J } \P{i\in\J}}{\sum_{\mathcal{A} = \J, \J^{\complement}}\P{i\in\bigcap_{l = 1}^{h} \Thi_l, i\in\bigcap_{l=1}^{m} \Thii_l^{\complement}| i\in\mathcal{A}}\P{i\in\mathcal{A}}} \nonumber \\
    & \overset{(a)}{=} \frac{\P{i\in\Thi|i\in\Ti}^h \P{i\in\Thi^{\complement}|i\in\Ti }^m \P{i\in\J}}{\sum_{\mathcal{A} = \J, \J^{\complement}}\P{i\in\Thi|i\in\mathcal{A}}^h\P{i\in\Thi^{\complement}|i\in\mathcal{A}}^m\P{i\in\mathcal{A}}} \nonumber \\
    & \overset{(b)}{=} \frac{(1-\epsilon)^h \epsilon^m \frac{J}{\cancel{N}}}{(1-\epsilon)^h \epsilon^m \frac{J}{\cancel{N}} + (\frac{T}{N-T}\epsilon)^h (1 - \frac{T}{N-T}\epsilon)^m \frac{N-J}{\cancel{N}}}
  \end{align}
  In \ensuremath{(a)} Lemma~\ref{lemma:indep} is applied, and lastly for \ensuremath{(b)}, Definition~\ref{ass} is used.
\end{IEEEproof}

\begin{IEEEproof}[Proof of Proposition~\ref{prop:common_cons_ineq}]
  This proposition states that he following inequality holds
  \begin{align}
    \P{i\in\J | i\in\bigcap_{l = 1}^{h}\Thi_l, i\in\bigcap_{l = 1}^{m}\Thii_l^{\complement}} \nonumber \\
    \geq \P{i\in\J | i\in\bigcap_{l=1}^{h-1}\Thi_l, i\in\bigcap_{l=1}^{m+1}\Thii_l^{\complement}} \label{proof:prop:common_cons_ineq}
  \end{align}
  
  Using Proposition~\ref{prop:common_cons_l}, the LHS of \eqref{proof:prop:common_cons_ineq} is:
  \begin{align}
    & \P{i\in\J | i\in\bigcap_{l=1}^{h}\Thi_l, i\in\bigcap_{l=1}^{m}\Thii_l^{\complement}} \nonumber \\
    & \phantom{=}= \frac{(1-\epsilon)^h \epsilon^m J}{(1-\epsilon)^h \epsilon^m J + (\frac{T}{N-T}\epsilon)^h (1 - \frac{T}{N-T}\epsilon)^m (N-J)}. \label{eqn:lhs1}
  \end{align}
  Similarly, the RHS of \eqref{proof:prop:common_cons_ineq} is
\begin{align}
  & \P{i\in\J | i\in\bigcap_{l=1}^{h-1}\Thi_l, i\in\bigcap_{l=1}^{m+1}\Thii_l^{\complement}} \nonumber \\
  & \phantom{=}= \frac{(1-\epsilon)^{h\!-\!1} \epsilon^{m\!+\!1} J}{(1\!-\!\epsilon)^{h\!-\!1} \epsilon^{m\!+\!1} J + (\frac{T}{N-T}\epsilon)^{h\!-\!1} (1\!-\!\frac{T}{N-T}\epsilon)^{m\!+\!1} (N\!-\!J)}. \label{eqn:rhs1}
  \end{align}
  By plugging the \eqref{eqn:lhs1} and \eqref{eqn:rhs1} into the inequality \eqref{proof:prop:common_cons_ineq} we get:
  \begin{align}
    & \frac{(1-\epsilon)^{\cancel{h}} \cancel{\epsilon^m} \cancel{J}}{(1-\epsilon)^h \epsilon^m J + (\frac{T}{N-T}\epsilon)^h (1 - \frac{T}{N-T}\epsilon)^m (N-J)} \nonumber \\
    & \geq \nonumber \\
    & \frac{\cancel{(1-\epsilon)^{h\!-\!1}} \epsilon^{\,\,\cancel{m\!+\!1}} \cancel{J}}{(1\!-\!\epsilon)^{h\!-\!1} \epsilon^{m\!+\!1} J + (\frac{T}{N-T}\epsilon)^{h\!-\!1} (1\!-\!\frac{T}{N-T}\epsilon)^{m\!+\!1} (N\!-\!J)} \nonumber
  \end{align}
  Multiplying the denominators gives
  \begin{align}
    & (1-\epsilon) \left( (1\!-\!\epsilon)^{h\!-\!1} \epsilon^{m\!+\!1} J \right. \nonumber \\
    & \phantom{(1-\epsilon)} \left. + (\frac{T}{N-T}\epsilon)^{h\!-\!1} (1\!-\!\frac{T}{N-T}\epsilon)^{m\!+\!1} (N\!-\!J) \right) \nonumber \\
    & \geq \nonumber  \\
    & \epsilon \left( (1-\epsilon)^h \epsilon^m J + (\frac{T}{N-T}\epsilon)^h (1 - \frac{T}{N-T}\epsilon)^m (N-J) \right),\nonumber
  \end{align}
  which by simplifying gives
  \begin{align}
    & \cancel{(1\!-\!\epsilon)^{h} \epsilon^{m\!+\!1} J} \nonumber \\
    & \phantom{=} + (1-\epsilon)(\frac{T}{N-T}\epsilon)^{h\!-\!1} (1\!-\!\frac{T}{N-T}\epsilon)^{m\!+\!1} (N\!-\!J) \nonumber \\
    & \geq \nonumber  \\
    & \cancel{(1-\epsilon)^h \epsilon^{m+1} J} + \epsilon(\frac{T}{N-T}\epsilon)^h (1 - \frac{T}{N-T}\epsilon)^m (N-J).\nonumber
  \end{align}
  Further simplifications give
  \begin{align}
    & (1-\epsilon)\cancel{(\frac{T}{N-T}\epsilon)^{h\!-\!1}} (1\!-\!\frac{T}{N-T}\epsilon)^{\cancel{m\!+\!1}} \cancel{(N\!-\!J)} \nonumber \\
    & \geq \nonumber  \\
    & \epsilon(\frac{T}{N-T}\epsilon)^{\cancel{h}} \cancel{(1 - \frac{T}{N-T}\epsilon)^m} \cancel{(N-J)}.\nonumber
  \end{align}

  Thus, we arrive at
  \begin{align}
    (1-\frac{T}{N-T}\epsilon) (1-\epsilon) \geq \frac{T}{N-T}\epsilon^2, \nonumber
  \end{align}
  which can be simplified to
  \begin{align}
     1 - \frac{N}{N-T}\epsilon + \cancel{\frac{T}{N-T}\epsilon^2} \geq \cancel{\frac{T}{N-T}\epsilon^2}. \nonumber
  \end{align}
  This is in turn equivalent to
  \begin{align}
    \frac{N}{N-T}\epsilon \leq 1, \nonumber
  \end{align}
  where the expression reaches its maximum at $\epsilon_{\max} = \frac{N-T}{N}$
  \begin{align}
    \frac{N}{N-T}\epsilon \leq \frac{N}{N-T}\frac{N-T}{N} = 1. \nonumber
  \end{align}
  Thus, we conclude that the sought inequality \eqref{eqn:common_cons_ineq} holds true.
\end{IEEEproof}

\subsection{Proof of the Results for \consensus}\label{app:consensus}
We now prove Proposition~\ref{cons:prop:pi}, Proposition~\ref{cons:prop:pic} and Remark~\ref{prop:ineq_example}, which are based on the mixed support-set model.

\begin{IEEEproof}[Proof of Proposition~\ref{cons:prop:pi}]
  We first notice that the problem can be split into two parts,
  \begin{align}
    &   \P{i\in\Ti | i\in\Thi, i\in\bigcap_{l = 1}^{h}\Thii_l, i\in\bigcap_{l = 1}^{m}\Thiii_l^{\complement}} \nonumber \\
    & = \P{i\in\J | i\in\Thi, i\in\bigcap_{l = 1}^{h}\Thii, i\in\bigcap_{l = 1}^{m}\Thiii_l^{\complement}} \label{eqn:pi_j_thi_thii_thiii} \\
    & \phantom{=} + \P{i\in\Ii | i\in\Thi, i\in\bigcap_{l = 1}^{h}\Thii_l, i\in\bigcap_{l = 1}^{m}\Thiii_l^{\complement}}, \label{eqn:pi_i_thi_thii_thiii}
  \end{align}
  where we consider each part separately.
  \begin{itemize}
  \item First we study \eqref{eqn:pi_j_thi_thii_thiii}
    \begin{align}
      & \P{i\in\J | i\in\Thi, i\in\bigcap_{l = 1}^{h}\Thii_l, i\in\bigcap_{l = 1}^{m}\Thiii_l^{\complement}} \nonumber \\
      & = \frac{\P{i\in\Thi, i\in\bigcap_{l = 1}^{h}\Thii_l, i\in\bigcap_{l = 1}^{m}\Thiii_l^{\complement} | i\in\J} \P{i\in\J}}{\P{i\in\Thi, i\in\bigcap_{l = 1}^{h}\Thii_l, i\in\bigcap_{l = 1}^{m}\Thiii_l^{\complement}}}. \label{eqn:pi_part1}
    \end{align}
    We now consider each probability in \eqref{eqn:pi_part1} separately, beginning with
    \begin{align}
      & \P{i\in\Thi, i\in\bigcap_{l = 1}^{h}\Thii_l, i\in\bigcap_{l = 1}^{m}\Thiii_l^{\complement} | i\in\J} \nonumber \\
      & = \P{i\in\bigcap_{l = 1}^{h+1}\Thi_l, i\in\bigcap_{l = 1}^{m}\Thii_l^{\complement} | i\in\J} \nonumber \\
      & \overset{(a)}{=} \P{i\in\Thi|i\in\Ti}^{h+1} \P{i\in\Thi^{\complement}|i\in\Ti}^{m} \nonumber \\
      & \overset{(b)}{=} (1-\epsilon)^{h+1} \epsilon^{m}. \label{eqn:pi_part2}
    \end{align}
    Here, Lemma~\ref{lemma:indep} was used in \ensuremath{(a)} and Definition~\ref{ass} was used in \ensuremath{(b)}. We have from the uniformity of the support-sets that
    \begin{align}
      \P{i\in\J} = \frac{J}{N}.
    \end{align}
    Finally we have
    \begin{align}
      & \P{i\in\Thi, i\in\bigcap_{l = 1}^{h}\Thii_l, i\in\bigcap_{l = 1}^{m}\Thiii_l^{\complement}} \nonumber \\
      & = \P{i\in\bigcap_{l = 1}^{h+1}\Thi_l, i\in\bigcap_{l = 1}^{m}\Thii_l^{\complement}} \nonumber \\
      & \overset{(a)}{=} (1-\epsilon)^{h+1} \epsilon^{m} \frac{J}{N} \nonumber \\
      & \phantom{=} + (h+1) (1-\epsilon) (\frac{T}{N-T}\epsilon)^{h}(1-\frac{T}{N-T}\epsilon)^m \frac{I}{N} \nonumber \\
      & \phantom{=} + m \epsilon (\frac{T}{N-T}\epsilon)^{h+1} (1-\frac{T}{N-T}\epsilon)^{m-1} \frac{I}{N} \nonumber \\
      & \phantom{=} + (\frac{T}{N-T}\epsilon)^{h+1} (1-\frac{T}{N-T}\epsilon)^{m} \frac{N-J-(m+h+1)I}{N}, \label{eqn:pi_part3}
    \end{align}
    where \ensuremath{(a)} is achieved by Lemma~\ref{lemma:joint_prob}.

  \item We now study \eqref{eqn:pi_i_thi_thii_thiii}
    \begin{align}
      & \P{i\in\Ii | i\in\Thi, i\in\bigcap_{l = 1}^{h}\Thii_l, i\in\bigcap_{l = 1}^{m}\Thiii_l^{\complement}} \nonumber \\
      & = \frac{\P{i\in\Thi, i\in\bigcap_{l = 1}^{h}\Thii_l, i\in\bigcap_{l = 1}^{m}\Thiii_l^{\complement} | i\in\Ii} \P{i\in\Ii}}{\P{i\in\Thi, i\in\bigcap_{l = 1}^{h}\Thii_l, i\in\bigcap_{l = 1}^{m}\Thiii_l^{\complement}}}. \label{eqn:pi_part4}
    \end{align}
    We now consider each probability in \eqref{eqn:pi_part4} separately, beginning with
    \begin{align}
      & \P{i\in\Thi, i\in\bigcap_{l = 1}^{h}\Thii_l, i\in\bigcap_{l = 1}^{m}\Thiii_l^{\complement} | i\in\Ii} \nonumber \\
      & \overset{(a)}{=} \P{i\in\Thi|i\in\Ti} \P{i\in\Thi|i\in\Ti^{\complement}}^{h} \P{i\in\Thi^{\complement}|i\in\Ti^{\complement}}^{m} \nonumber \\
      & \overset{(b)}{=} (1-\epsilon) (\frac{T}{N-T}\epsilon)^{h} (1 - (\frac{T}{N-T}\epsilon))^{m}, \nonumber
    \end{align}
    where we just as for \eqref{eqn:pi_part2}, used Lemma~\ref{lemma:indep} for \ensuremath{(a)} and Definition~\ref{ass} for \ensuremath{(b)}. We have from the uniformity of support-sets that,
    \begin{align}
      \P{i\in\Ii} = \frac{I}{N}.
    \end{align}
    Finally we notice for the third probability that the denominator is identical to \eqref{eqn:pi_part3}. Now plugging the parts together gives \eqref{eqn:pi_phi_phii_phiiic}.
  \end{itemize}
\end{IEEEproof}

\begin{IEEEproof}[Proof of Proposition~\ref{cons:prop:pic}]
  This proof is similar to the proof of Proposition~\ref{cons:prop:pi}. First, split the problem into two parts,
  \begin{align}
    &   \P{i\in\Ti | i\in\Thi^{\complement}, i\in\bigcap_{l = 1}^{h}\Thii_l, i\in\bigcap_{l = 1}^{m}\Thiii_l^{\complement}} \nonumber \\
    & = \P{i\in\J | i\in\Thi^{\complement}, i\in\bigcap_{l = 1}^{h}\Thii, i\in\bigcap_{l = 1}^{m}\Thiii_l^{\complement}} \label{eqn:pic_j_thi_thii_thiii} \\
    & \phantom{=} + \P{i\in\Ii | i\in\Thi^{\complement}, i\in\bigcap_{l = 1}^{h}\Thii_l, i\in\bigcap_{l = 1}^{m}\Thiii_l^{\complement}}, \label{eqn:pic_i_thi_thii_thiii}
  \end{align}
  and we study each part separately.
  \begin{itemize}
  \item First study \eqref{eqn:pic_j_thi_thii_thiii}
    \begin{align}
      & \P{i\in\J | i\in\Thi^{\complement}, i\in\bigcap_{l = 1}^{h}\Thii_l, i\in\bigcap_{l = 1}^{m}\Thiii_l^{\complement}} \nonumber \\
      & = \frac{\P{i\in\Thi^{\complement}, i\in\bigcap_{l = 1}^{h}\Thii_l, i\in\bigcap_{l = 1}^{m}\Thiii_l^{\complement} | i\in\J} \P{i\in\J}}{\P{i\in\Thi^{\complement}, i\in\bigcap_{l = 1}^{h}\Thii_l, i\in\bigcap_{l = 1}^{m}\Thiii_l^{\complement}}}. \label{eqn:pic_part1}
    \end{align}
    This was achieved using Bayes' rule. We now consider each probability in \eqref{eqn:pic_part1} separately, beginning with
    \begin{align}
      & \P{i\in\Thi^{\complement}, i\in\bigcap_{l = 1}^{h}\Thii_l, i\in\bigcap_{l = 1}^{m}\Thiii_l^{\complement} | i\in\J} \\
      & = \P{i\in\bigcap_{l = 1}^{h}\Thii_l, i\in\bigcap_{l = 1}^{m+1}\Thiii_l^{\complement} | i\in\J} \\
      & \overset{(a)}{=} \P{i\in\Thi| i\in\Ti}^h \P{i\in\Thi^{\complement} | i\in\Ti}^{m+1} \\
      & \overset{(b)}{=} (1-\epsilon)^h\epsilon^{m+1}
    \end{align}
    Here, Lemma~\ref{lemma:indep} was used in \ensuremath{(a)} and Definition~\ref{ass} was used in \ensuremath{(b)}. We have from the uniformity of the support-sets that
    \begin{align}
      \P{i\in\J} = \frac{J}{N}.
    \end{align}
    Finally we have
    \begin{align}
      & \P{i\in\Thi^{\complement}, i\in\bigcap_{l = 1}^{h}\Thii_l, i\in\bigcap_{l = 1}^{m}\Thiii_l^{\complement}} \\
      & = \P{i\in\bigcap_{l = 1}^{h}\Thii_l, i\in\bigcap_{l = 1}^{m+1}\Thiii_l^{\complement}} \\
      & \overset{(a)}{=} (1-\epsilon)^{h} \epsilon^{m+1} \frac{J}{N} \nonumber \\
      & \phantom{=} + h (1-\epsilon) (\frac{T}{N-T}\epsilon)^{h-1}(1-\frac{T}{N-T}\epsilon)^{m+1} \frac{I}{N} \nonumber \\
      & \phantom{=} + (m+1) \epsilon (\frac{T}{N-T}\epsilon)^{h} (1-\frac{T}{N-T}\epsilon)^{m} \frac{I}{N} \nonumber\\
      & \phantom{=} + (\frac{T}{N-T}\epsilon)^{h} (1-\frac{T}{N-T}\epsilon)^{m+1} \frac{N - J -(m+h+1)I}{N}, \label{eqn:pic_part3}
    \end{align}
    where \ensuremath{(a)} is obtained by Lemma~\ref{lemma:joint_prob}.

  \item For \eqref{eqn:pic_i_thi_thii_thiii} we have that
    \begin{align}
      & \P{i\in\Ii | i\in\Thi^{\complement}, i\in\bigcap_{l = 1}^{h}\Thii_l, i\in\bigcap_{l = 1}^{m}\Thiii_l^{\complement}} \nonumber \\
      & = \frac{\P{i\in\Thi^{\complement}, i\in\bigcap_{l = 1}^{h}\Thii_l, i\in\bigcap_{l = 1}^{m}\Thiii_l^{\complement} | i\in\Ii} \P{i\in\Ii}}{\P{i\in\Thi^{\complement}, i\in\bigcap_{l = 1}^{h}\Thii_l, i\in\bigcap_{l = 1}^{m}\Thiii_l^{\complement}}}, \label{eqn:pic_part2}
    \end{align}
    which is achieved with Bayes' rule. We now consider each probability in \eqref{eqn:pic_part2} separately, beginning with the first probability
    \begin{align}
      & \P{i\in\Thi^{\complement}, i\in\bigcap_{l = 1}^{h}\Thii_l, i\in\bigcap_{l = 1}^{m}\Thiii_l^{\complement} | i\in\Ii} \nonumber \\
      & = \P{i\in\bigcap_{l = 1}^{h}\Thii_l, i\in\bigcap_{l = 1}^{m+1}\Thiii_l^{\complement} | i\in\Iiii_{{m+1}}} \nonumber \\
      & \overset{(a)}{=} \P{i\in\Thi^{\complement}| i\in\Ti} \P{i\in\Thii| i\in\Tii^{\complement}}^h \P{\Thiii^{\complement} | i\in\Tiii^{\complement}}^m \nonumber \\
      & \overset{(b)}{=} \epsilon (\frac{T}{N-T}\epsilon)^h (1-\frac{T}{N-T}\epsilon)^m, \nonumber
    \end{align}
    where we used Lemma~\ref{lemma:indep} for \ensuremath{(a)} and Definition~\ref{ass} for \ensuremath{(b)}. We have from the uniformity of support-sets that,
    \begin{align}
      \P{i\in\Ii} = \frac{I}{N}. \nonumber
    \end{align}
    Finally we notice for the third probability that the denominator is identical to \eqref{eqn:pic_part3}. Now plugging all the parts together gives \eqref{eqn:pi_phic_phii_phiiic}.
  \end{itemize}
\end{IEEEproof}

\begin{IEEEproof}[Proof of Remark~\ref{prop:ineq_example}]
  We first study \eqref{eqn:jh} and notice that any index in $i\in\Jh_{\p}$ fulfills one of the following: $i\in(\Thi \cap \Thii \cap \Thiii)$, or $i\in(\Thi \cap \Thii \cap \Thiii^{\complement})$, or $i\in(\Thi^{\complement} \cap \Thii \cap \Thiii)$. Thus, we will show the remark by proving each of the following inequalities:
  \begin{align}
    \P{i\in\Ti | i\in (\Thi \cap \Thii \cap \Thiii)} & \geq \P{i\in\Ti | i\in\Thi}, \label{prop:ineq_example_1} \\
    \P{i\in\Ti | i\in (\Thi \cap \Thii \cap \Thiii^{\complement})} & \geq \P{i\in\Ti | i\in\Thi}, \label{prop:ineq_example_2} \\
    \P{i\in\Ti | i\in (\Thi^{\complement} \cap \Thii \cap \Thiii)} & \geq \P{i\in\Ti | i\in\Thi}. \label{prop:ineq_example_3}
  \end{align}
  First, recall from Proposition~\ref{prop:disconnect} and \eqref{ass2} that
  \begin{align}
    \P{i\in\Ti | i\in\Thi} = 1 - \epsilon. \label{eqn:ex1_1}
  \end{align}
  \begin{itemize}

  \item We now consider \eqref{prop:ineq_example_1}. By plugging $N = 1000$, $T = 20$, $J = 15$, $I = 5$, $m = 0$ and $h = 2$ into Proposition~\ref{cons:prop:pi}, we obtain
    \begin{align}
      & \P{i\in\Ti | i\in (\Thi \cap \Thii \cap \Thiii)} \label{eqn:ex1_2} \\
      & = \frac{49(\epsilon - 1)(7204\epsilon^2 - 14406\epsilon + 7203)}{352900\epsilon^3 - 1058988\epsilon^2 + 1058841\epsilon - 352947}. \nonumber
    \end{align}
    We multiply the denominator of \eqref{eqn:ex1_2} to \eqref{eqn:ex1_1} get the following inequality
    \begin{align}
      & 49(\epsilon - 1)(7204\epsilon^2 - 14406\epsilon + 7203) \nonumber \\
      & \geq (1-\epsilon)(352900\epsilon^3 - 1058988\epsilon^2 + 1058841\epsilon - 352947), \nonumber
    \end{align}
    which equivalently can be simplified to
    \begin{align}
      0 \geq (\epsilon(50\epsilon - 49)(7058\epsilon - 7203)(\epsilon - 1))/23529800. \label{eqn:zeros1}
    \end{align}
    The roots to the polynomial of \eqref{eqn:zeros1} are: $\epsilon_1 = 0$, $\epsilon_2 = \frac{49}{50} = 0.98$, $\epsilon_3 = \frac{7203}{7058} = 1.0205...$ and $\epsilon_4 = 1$. Thus, the interesting region is $\epsilon \in [\epsilon_1, \epsilon_2]$, for which the inequality \eqref{eqn:zeros1} holds.
    %{\color{red} Find roots}

  \item We now consider \eqref{prop:ineq_example_2}. By plugging $N = 1000$, $T = 20$, $J = 15$, $I = 5$, $m = 1$ and $h = 1$ into Proposition~\ref{cons:prop:pi}, we obtain
    \begin{align}
      & \P{i\in\Ti | i\in (\Thi \cap \Thii \cap \Thiii^{\complement})} \nonumber \\
      & = \frac{196\epsilon(1801\epsilon^2 - 3614\epsilon + 1813)}{\epsilon(352900\epsilon^2 - 701288\epsilon + 357749)}. \label{eqn:ex3_1}
    \end{align}
    Observe that also here, $\epsilon = 0$ is undefined. We multiply the denominator of \eqref{eqn:ex3_1} to \eqref{eqn:ex1_1} and get the following inequality
    \begin{align}
      &  196\epsilon(1801\epsilon^2 - 3614\epsilon + 1813) \nonumber \\
      & \geq \epsilon(1-\epsilon)(352900\epsilon^2 - 701288\epsilon + 357749), \nonumber
    \end{align}
    which can be simplified to
    \begin{align}
      0 \geq -(\epsilon(50\epsilon - 49)(7058\epsilon - 49)(\epsilon - 1))/23529800. \label{eqn:zeros2}
    \end{align}
    The roots to the polynomial of \eqref{eqn:zeros2} are: $\epsilon_1 = 0$, $\epsilon_2 = \frac{49}{50} = 0.98$, $\epsilon_3 = \frac{49}{7058} = 0.0069...$ and $\epsilon_4 = 1$. Thus, the interesting region is $\epsilon \in [\epsilon_3, \epsilon_4]$, for which the inequality \eqref{eqn:zeros2} holds.
%Thus, the interesting region is between $\epsilon_3$ and $\epsilon_2$; in which by plugging in any value in between, the inequality \eqref{eqn:zeros2} turns out to be fulfilled.
    
  \item We now consider \eqref{prop:ineq_example_3}. By plugging $N = 1000$, $T = 20$, $J = 15$, $I = 5$, $m = 0$ and $h = 2$ into Proposition~\ref{cons:prop:pic}, we obtain
    \begin{align}
      & \P{i\in\Ti | i\in (\Thi^{\complement} \cap \Thii \cap \Thiii)} \nonumber \\
      & = \frac{49\epsilon(7204\epsilon^2 - 14406\epsilon + 7203)}{\epsilon(352900\epsilon^2 - 701288\epsilon + 357749)}. \label{eqn:ex2_1}
    \end{align}
    Observe that in this expression, $\epsilon = 0$ is undefined. This is naturally true\footnote{If all algorithms are perfect, the cut $(\Thi \cap \Thii \cap \Thiii^{\complement}) = \emptyset$.} and follows directly from Lemma~\ref{lemma:joint_prob}. We multiply the denominator of \eqref{eqn:ex2_1} to \eqref{eqn:ex1_1} and get the following inequality
    \begin{align}
      & 49\epsilon(7204\epsilon^2 - 14406\epsilon + 7203) \nonumber \\
      & \geq \epsilon(1-\epsilon)(352900\epsilon^2 - 701288\epsilon + 357749), \nonumber
    \end{align}
    which can be simplified to
    \begin{align}
      0 \geq -(\epsilon(50\epsilon - 49)(7058\epsilon^2 - 7107\epsilon + 98))/23529800. \label{eqn:zeros3}
    \end{align}
    The roots to the polynomial of \eqref{eqn:zeros3} are: $\epsilon_1 = 0$, $\epsilon_2 = \frac{49}{50} = 0.98$, $\epsilon_3 = \frac{7107}{2\cdot 7058} + \sqrt{\left(\frac{7107}{2\cdot 7058}\right)^2 - \frac{98}{7058}} = 0.9930...$ and $\epsilon_4 = \frac{7107}{2\cdot 7058} - \sqrt{\left(\frac{7107}{2\cdot 7058}\right)^2 - \frac{98}{7058}} = 0.0140...$. Thus, the interesting region is $\epsilon \in [\epsilon_4, \epsilon_3]$, for which the inequality \eqref{eqn:zeros3} holds.
%Thus, the interesting region is between $\epsilon_3$ and $\epsilon_4$; in which by plugging in any value in between, the inequality \eqref{eqn:zeros3} turns out to be fulfilled.

%    The roots to the polynomial of \eqref{eqn:zeros2} are: $\epsilon_1 = 0$, $\epsilon_2 = \frac{49}{50} = 0.98$, $\epsilon_3 = \frac{835}{2\cdot 786} + \sqrt{(\frac{835}{2\cdot 786})^2 - \frac{98}{786}} = 0.9280...$ and $\epsilon_4 = 0.1344...$. Thus, the interesting region is between $\epsilon_3$ and $\epsilon_2$; which by plugging in to \eqref{eqn:zeros2} turns out to be fulfilled.
%    {\color{red} Find roots}

  \end{itemize}
  From the above calculations, we find the interesting region is the region that lies between $\epsilon_4 \geq 0.0140$ for \eqref{eqn:zeros3} and $\epsilon_{\max} = \frac{N-T}{N} = 0.98$. Since all the inequalities \eqref{eqn:zeros1}, \eqref{eqn:zeros2} or \eqref{eqn:zeros3} hold true in this region (directly verified by plugging in any $0.0140 \leq \epsilon \leq 0.98$), we conclude the proof.  
\end{IEEEproof}

\ifCLASSOPTIONcaptionsoff
  \newpage
\fi



\bibliographystyle{IEEEtran}
\bibliography{references/IEEEfull,references/myconffull,references/compressed_sensing}

\begin{thebibliography}{10}
\providecommand{\url}[1]{#1}
\csname url@samestyle\endcsname
\providecommand{\newblock}{\relax}
\providecommand{\bibinfo}[2]{#2}
\providecommand{\BIBentrySTDinterwordspacing}{\spaceskip=0pt\relax}
\providecommand{\BIBentryALTinterwordstretchfactor}{4}
\providecommand{\BIBentryALTinterwordspacing}{\spaceskip=\fontdimen2\font plus
\BIBentryALTinterwordstretchfactor\fontdimen3\font minus
  \fontdimen4\font\relax}
\providecommand{\BIBforeignlanguage}[2]{{%
\expandafter\ifx\csname l@#1\endcsname\relax
\typeout{** WARNING: IEEEtran.bst: No hyphenation pattern has been}%
\typeout{** loaded for the language `#1'. Using the pattern for}%
\typeout{** the default language instead.}%
\else
\language=\csname l@#1\endcsname
\fi
#2}}
\providecommand{\BIBdecl}{\relax}
\BIBdecl

\bibitem{Donoho:compressed_sensing}
D.~Donoho, ``Compressed sensing,'' \emph{{IEEE} Transactions on Information
  Theory}, vol.~52, pp. 1289--1306, Apr. 2006.

\bibitem{Candes:stable_signal_recovery}
E.~J. Cand\`{e}s, J.~Romberg, and T.~Tao, ``Stable signal recovery from
  incomplete and inaccurate measurements,'' \emph{Communications on Pure and
  Applied Mathematics}, vol.~59, pp. 1207--1223, Aug. 2006.

\bibitem{Mota:distributed_basis_pursuit}
J.~Mota, J.~Xavier, P.~Aguiar, and M.~Puschel, ``Distributed basis pursuit,''
  \emph{{IEEE} Transactions on Signal Processing}, vol.~60, pp. 1942--1956,
  Apr. 2012.

\bibitem{Bazerque:distributed_spectrum_sensing}
J.~Bazerque and G.~Giannakis, ``Distributed spectrum sensing for cognitive
  radio networks by exploiting sparsity,'' \emph{{IEEE} Transactions on Signal
  Processing}, vol.~58, pp. 1847--1862, Mar. 2010.

\bibitem{Ji:bayesian_compressive_sensing}
S.~Ji, Y.~Xue, and L.~Carin, ``Bayesian compressive sensing,'' \emph{{IEEE}
  Transactions on Signal Processing}, vol.~56, pp. 2346--2356, Jun. 2008.

\bibitem{Donoho:message_passing_for_cs}
D.~Donoho, A.~Maleki, and A.~Montanari, ``Message-passing algorithms for
  compressed sensing,'' \emph{Proceedings of the National Academy of Sciences},
  vol. 106, pp. 18\,914--18\,919, Oct. 2009.

\bibitem{Mallat:matching_pursuit_with_time_frequency_dictionaries}
S.~Mallat and Z.~Zhang, ``Matching pursuits with time-frequency dictionaries,''
  \emph{{IEEE} Transactions on Signal Processing}, vol.~41, pp. 3397--3415,
  Dec. 1993.

\bibitem{Tropp:signal_recovery}
J.~Tropp and A.~Gilbert, ``Signal recovery from random measurements via
  orthogonal matching pursuit,'' \emph{{IEEE} Transactions on Information
  Theory}, vol.~53, pp. 4655--4666, Dec. 2007.

\bibitem{Needell:cosamp}
D.~Needell and J.~A. Tropp, ``Cosamp: Iterative signal recovery from incomplete
  and inaccurate samples,'' \emph{Applied and Computational Harmonic Analysis},
  vol.~26, pp. 301--321, Apr. 2009.

\bibitem{Dai:subspace_pursuit}
W.~Dai and O.~Milenkovic, ``Subspace pursuit for compressive sensing signal
  reconstruction,'' \emph{{IEEE} Transactions on Information Theory}, vol.~55,
  pp. 2230--2249, May 2009.

\bibitem{Donoho:sparse_solution_of_underdetermined_systems_stomp}
D.~Donoho, Y.~Tsaig, I.~Drori, and J.-L. Starck, ``Sparse solution of
  underdetermined systems of linear equations by stagewise orthogonal matching
  pursuit,'' \emph{{IEEE} Transactions on Information Theory}, vol.~58, pp.
  1094--1121, Feb. 2012.

\bibitem{Chatterjee:projection_based_look_ahead}
S.~Chatterjee, D.~Sundman, M.~Vehkaper{\"a}, and M.Skoglund, ``Projection-based
  and look-ahead strategies for atom selection,'' \emph{{IEEE} Transactions on
  Signal Processing}, vol.~60, pp. 634--647, Feb. 2012.

\bibitem{Sundman:frogs}
D.~Sundman, S.~Chatterjee, and M.~Skoglund, ``{FROGS}: A serial reversible
  greedy search algorithm,'' in \emph{IEEE Swedish Communication Technologies
  Workshop (Swe-CTW)}, Lund, Sweden, Mar. 2012, pp. 40--45.

\bibitem{Needell:signal_recovery_from_incomplete_and_inaccurate_measurements_via_romp}
D.~Needell and R.~Vershynin, ``Signal recovery from incomplete and inaccurate
  measurements via regularized orthogonal matching pursuit,'' \emph{{IEEE}
  Journal of Selected Topics in Signal Processing}, vol.~4, pp. 310--316, Mar.
  2010.

\bibitem{Sundman:look_ahead_parallel_pursuit}
D.~Sundman, S.~Chatterjee, and M.~Skoglund, ``Look ahead parallel pursuit,'' in
  \emph{IEEE Swedish Communication Technologies Workshop (Swe-CTW)}, Stockholm,
  Sweden, Oct. 2011, pp. 114--117.

\bibitem{Candes:restricted_isometry_property}
E.~J. Cand\`{e}s, ``The restricted isometry property and its implications for
  compressed sensing,'' \emph{Comptes Rendus Mathematique}, vol. 346, pp.
  589--592, May 2008.

\bibitem{Tropp:simultaneous_sparse_approx_part1}
J.~Tropp, A.~Gilbert, and M.~Strauss, ``Algorithms for simultaneous sparse
  approximation. part i: Greedy pursuit,'' \emph{Signal Processing}, vol.~86,
  pp. 572--588, Mar. 2006.

\bibitem{Rakotomamonjy:surveying}
A.~Rakotomamonjy, ``Surveying and comparing simultaneous sparse approximation
  (or group-lasso) algorithms,'' \emph{Signal Processing}, vol.~91, pp.
  1505--1526, Jul. 2011.

\bibitem{Leviatan:simultaneous}
D.~Leviatan and V.~Temlyakov, ``Simultaneous approximation by greedy
  algorithms,'' \emph{Advances in Computational Mathematics}, vol.~25, pp.
  73--90, Jul. 2006.

\bibitem{Cotter:sparse_solutions}
S.~Cotter, B.~Rao, K.~Engan, and K.~Kreutz-Delgado., ``Sparse solutions to
  linear inverse problems with multiple measurement vectors,'' \emph{{IEEE}
  Transactions on Signal Processing}, vol.~53, pp. 2477--2488, Jul. 2005.

\bibitem{Chen:theoretical_results_on_sparse}
J.~Chen and X.~Huo, ``Theoretical results on sparse representations of
  multiple-measurement vectors,'' \emph{{IEEE} Transactions on Signal
  Processing}, vol.~54, pp. 4634--4643, Dec. 2006.

\bibitem{Sundman:greedy_pursuit_for_jointly}
D.~Sundman, S.~Chatterjee, and M.~Skoglund, ``Greedy pursuits for compressed
  sensing of jointly sparse signals,'' in \emph{EURASIP European Signal
  Processing Conference (EUSIPCO)}, Barcelona, Spain, Aug. 2011, pp. 368--372.

\bibitem{Bazeraque:distributed_spectrum_sensing}
J.~Bazerque and G.~Giannakis, ``Distributed spectrum sensing for cognitive
  radio networks by exploiting sparsity,'' \emph{{IEEE} Transactions on Signal
  Processing}, vol.~58, pp. 1847--1862, Mar. 2010.

\bibitem{Feng:distributed_compressive_spectrum_sensing}
F.~Zeng, C.~Li, and Z.~Tian, ``Distributed compressive spectrum sensing in
  cooperative multihop cognitive networks,'' vol.~5, pp. 37--48, Feb. 2011.

\bibitem{Ling:decenteralized_support_detection}
Q.~Ling and T.~Zhi, ``Decentralized support detection of multiple measurement
  vectors with joint sparsity,'' in \emph{IEEE International Conference on
  Acoustics, Speech and Signal Processing (ICASSP)}, Prague, Czech Republic,
  May 2011, pp. 2996--2999.

\bibitem{Sundman:diprsp}
D.~Sundman, D.~Zachariah, and S.~Chatterjee, ``Distributed predictive subspace
  pursuit,'' in \emph{IEEE International Conference on Acoustics, Speech and
  Signal Processing (ICASSP)}, Vancouver, Canada, May 2013, pp. 4633--4637.

\bibitem{Sundman:a_greedy_pursuit_algorithm}
D.~Sundman, S.~Chatterjee, and M.~Skoglund, ``A greedy pursuit algorithm for
  distributed compressed sensing,'' in \emph{IEEE International Conference on
  Acoustics, Speech and Signal Processing (ICASSP)}, Kyoto, Japan, Mar. 2012,
  pp. 2729--2732.

\bibitem{Sundman:distributed_gp_algorithms}
------, ``Distributed greedy pursuit algorithms,'' \emph{Signal Processing},
  2014, accepted.

\bibitem{Sundman:dipp_arxiv}
------, ``{DIPP}: A distributed greedy algorithm based on democratic voting
  principles,'' \emph{CoRR}, vol. abs/1403.6974, May 2014.

\bibitem{Sundman:methods_for_dcs}
\BIBentryALTinterwordspacing
------, ``Methods for distributed compressed sensing,'' \emph{MDPI Journal of
  Sensor and Actuator Networks}, vol.~3, pp. 1--25, Dec. 2013. [Online].
  Available: \url{http://www.mdpi.com/2224-2708/3/1/1}
\BIBentrySTDinterwordspacing

\bibitem{Young:optimal_voting_rules}
P.~Young, ``Optimal voting rules,'' \emph{The Journal of Economic
  Perspectives}, vol.~9, pp. 51--64, Dec. 1995.

\bibitem{Ledyard:the_approximation_of_efficient}
J.~Ledyard and T.~Palfrey, ``The approximation of efficient public good
  mechanisms by simple voting schemes,'' \emph{Journal of Public Economics},
  vol.~83, pp. 153--171, Feb. 2002.

\bibitem{Duarte:distributed_compressed_sensing}
M.~Duarte, S.~Sarvotham, D.~Baron, M.~Wakin, and R.~Baraniuk, ``Distributed
  compressed sensing of jointly sparse signals,'' in \emph{Annual Asilomar
  Conference on Signals, Systems, and Computers}, Pacific Grove, California,
  Nov. 2005, pp. 1537--1541.

\end{thebibliography}
\end{document}